\documentclass{statsoc}

\usepackage[a4paper]{geometry}
\usepackage{graphicx}
\usepackage{amsmath}
\usepackage{natbib}
\usepackage{appendix}
\usepackage{url}
\usepackage{array}
\newcolumntype{L}[1]{>{\raggedright\let\newline\\\arraybackslash\hspace{0pt}}m{#1}}
\newcolumntype{C}[1]{>{\centering\let\newline\\\arraybackslash\hspace{0pt}}m{#1}}
\newcolumntype{R}[1]{>{\raggedleft\let\newline\\\arraybackslash\hspace{0pt}}m{#1}}
\usepackage{amsmath}
\newcommand{\bc}{\begin{center}}
\newcommand{\ec}{\end{center}}
\newcommand{\be}{\begin{enumerate}}
\newcommand{\ee}{\end{enumerate}}
\newcommand{\bi}{\begin{itemize}}
\newcommand{\ei}{\end{itemize}}
\newcommand{\bt}[1]{\begin{tabular}{#1}}
\newcommand{\et}{\end{tabular}}
\newcommand{\bd}{\begin{description}}
\newcommand{\ed}{\end{description}}
\newcommand{\bal}{\begin{align}}
\newcommand{\eal}{\end{align}}
\newcommand{\bals}{\begin{align*}}
\newcommand{\eals}{\end{align*}}

\newcommand{\bes}{\begin{eqnarray*}}
\newcommand{\ees}{\end{eqnarray*}}
\newcommand{\bs}{\begin{slide}{}}
\newcommand{\es}{\end{slide}}

\newcommand{\mbf}[1]{\mbox{\boldmath$#1$}}
\newcommand{\mat}[2]{\mbox{$\left[\begin{array}{#1}#2\end{array}\right]$}}

\newcommand{\bbmp}{\begin{boxedminipage}}
\newcommand{\ebmp}{\end{boxedminipage}}

\newcommand{\bprimary}{\mbox{$\hat{B}_{\textnormal{\tiny primary}}$}}
\newcommand{\wprimary}{\mbox{$\hat{W}_{\textnormal{\tiny primary}}$}}

\newcommand{\vsfulhat}{\mbox{$\hat{V}_{\textnormal{\tiny full, sensitivity}}$}}
\newcommand{\vpfulhat}{\mbox{$\hat{V}_{\textnormal{\tiny full, primary}}$}}

\newcommand{\vpmshat}{\mbox{$\hat{V}_{\textnormal{\tiny obs, primary}}$}}

\newcommand{\vpmihat}{\mbox{$\hat{V}_{\textnormal{\tiny MI, primary}}$}}

\newcommand{\vsmihat}{\mbox{$\hat{V}_{\textnormal{\tiny MI, sensitivity}}$}}
\newcommand{\bsen}{\mbox{$\hat{B}_{\textnormal{\tiny sensitivity}}$}}
\newcommand{\wsen}{\mbox{$\hat{W}_{\textnormal{\tiny sensitivity}}$}}
\newcommand{\vrubin}{\mbox{$\hat{V}_{\textnormal{\tiny Rubin's, MI}}$}}
\newcommand{\vanchat}{\mbox{$\hat{V}_{\textnormal{\tiny anchored}}$}}

\newcommand{\btheh}{\mbox{$\mbf{\hat{\theta}}$}} 

\newcommand{\bb}{\mbox{\bfseries b}}

\newcommand{\boldi}{\mbox{\bfseries i}}

\newcommand{\bA}{\mbox{\bfseries A}}

\newcommand{\bE}{\mbox{\bfseries E}}
\newcommand{\E}{\mbox{\bfseries E}}

\newcommand{\bL}{\mbox{\bfseries L}}

\newcommand{\bP}{\mbox{\bfseries P}}

\newcommand{\bR}{\mbox{\bfseries R}}

\newcommand{\bV}{\mbox{\bfseries V}}

\newcommand{\bY}{\mbox{\bfseries Y}}

\newcommand{\bzero}{\mbox{\bfseries 0}}

\newcommand{\fev}{\mbox{$\text{FEV}_1$}}

\newcommand{\bthe}{\mbox{$\mbf{\theta}$}}
\newcommand{\beps}{\mbox{$\mbf{\epsilon}$}}

\newcommand{\bbe}{\mbox{$\mbf{\beta}$}}
\newcommand{\bmu}{\mbox{$\mbf{\mu}$}}

\newcommand{\bbeh}{\mbox{$\hat{\bbe}$}}

\newcommand{\bSigma}{\mbox{$\mbf{\Sigma}$}}

\title[Information-Anchored Sensitivity Analysis]{Information-Anchored
  Sensitivity Analysis:\\ Theory and Application\\
}
\author{Suzie Cro}
\address{MRC Clinical Trials Unit at UCL, UK,  London School of Hygiene \& Tropical Medicine, UK and Imperial College London, UK.}
\email{s.cro@imperial.ac.uk}
\author{James R Carpenter}
\address{MRC Clinical Trials Unit at UCL, UK and London School of Hygiene \& Tropical Medicine, UK.}
\author[Cro, Carpenter \& Kenward]{Michael G Kenward}
\address{Ashkirk, Scotland.}

\begin{document}

\begin{abstract}
Analysis of longitudinal randomised controlled trials is frequently
complicated because patients deviate from the protocol. Where such
deviations are relevant for the estimand, we are typically required to
make an untestable assumption about post-deviation behaviour in order to
perform our primary analysis and estimate the treatment effect.
In such settings, it is now widely recognised that we should follow
this with sensitivity analyses to explore the robustness of our
inferences to alternative assumptions about post-deviation behaviour. 
Although there has been a lot of work on how to conduct such
sensitivity analyses, little attention has been given to the
appropriate loss of information due to missing data within sensitivity
analysis. We argue more attention needs to be given to this issue,
showing it is quite possible for sensitivity analysis to decrease and
increase the information about the treatment effect.
To address this critical issue, we introduce the concept of {\em
  information-anchored} sensitivity analysis. By this we mean
sensitivity analysis in which the proportion of information about the treatment estimate lost
due to missing data is the same
as the proportion of information about the treatment estimate lost
due to missing data in the primary analysis. We argue this forms a
transparent, practical starting point for interpretation of
sensitivity analysis. We then derive results showing that, for
longitudinal continuous data, a broad
class of controlled and reference-based sensitivity analyses performed by multiple imputation are
information-anchored. We illustrate the theory with simulations and an
analysis of a peer review trial, then discuss our work in the context of
other recent work in this area. 
Our results  give a  theoretical basis for the use of
controlled multiple imputation procedures for sensitivity analysis.

\end{abstract}


\section{Introduction}

\label{sec:intro}

The statistical analysis of longitudinal randomised clinical
trials is frequently complicated because patients deviate from the
trial protocol. Such deviations are increasingly referred to as
  inter-current events. For example, patients might withdraw from trial treatment, switch treatment,  receive additional rescue therapy or simply become lost to follow-up. Post-deviation, such patients' data (if available)
will often no longer be directly relevant for the primary
estimand. Consequently, such post-deviation data are often set as missing; any observed
post-deviation data can then inform the missing data assumptions. Nevertheless, however  the analysis is approached,  unverifiable assumptions
about aspects of the statistical distribution of the post-deviation data must  be made.

Recognising this, recent regulatory guidelines from the European Medicines Agency
\cite{EMA:2010} and a United States Food and Drug Administration mandated
panel report by the \cite{NRC:2010}
emphasise the  importance of conducting sensitivity analyses.
Further, the recent publication of the International Conference on Harmonisation of Technical Requirements for Registration of Pharmaceuticals for Human Use (ICH) E9 (R1) addendum on
estimands and sensitivity analysis
 in clinical trials (\citeyear{ICH:2017}) raises important issues about how such
 sensitivity analyses should be approached. It highlights how in any
 trial setting it is important first to define the estimand of
 interest. This will inform what data are missing and how such missing data
 should be handled in the primary analysis. Sensitivity analysis,
 which targets the same estimand, should subsequently be undertaken to
 address the robustness of inferences to the underlying
 assumptions, including those made for the missing data.

We propose splitting sensitivity analyses for missing data into two broad classes. In both classes, one or more alternative sets 
of assumptions (or scenarios) are postulated and the sensitivity of the conclusions to these alternative scenarios 
is to be assessed. In our first class, the primary analysis model is retained in
the sensitivity analysis. This enables the exclusive assessment of the impact of alternative missing data assumptions
on the primary outcome of interest. For example, for our sensitivity analysis we may impute missing data
under a missing not at random (MNAR) assumption, and fit the primary analysis
model to these imputed data. When performed by Multiple Imputation (MI), class-1 sensitivity analyses are therefore uncongenial, in the sense described by  \cite{Meng:1994} and \cite{Xianchao:2017}.
Conversely, in the second class, for each set of sensitivity
assumptions an appropriate analysis model is identified and
fitted. Hence, each such analysis model 
is consistent with  its assumptions, which is why the analysis models
generally change as we move from scenario to
scenario.

In the first class of sensitivity analyses, the assumptions of the
primary analysis model may be inconsistent to some degree with the data generating
mechanism postulated by the sensitivity analysis
assumption. Nevertheless, a strong advantage of such sensitivity analysis is the avoidance of full modelling under
various, potentially very complex, missing data assumptions. However,
when performing class-1 sensitivity analyses, the properties of an
 estimator under the primary analysis may change as we move to the
 sensitivity analysis. In particular, we will see that a sensible variance estimator
for the primary analysis may behave in an unexpected way under certain
sensitivity analysis scenarios, for example decreasing as the proportion of missing
values increases. In regulatory work,  particularly in class-1 sensitivity analyses,
it is therefore important to appreciate fully
the quantity and nature of any additional statistical information about the treatment
estimate that may arise in the sensitivity analysis, relative to the
\mbox{primary analysis.} 

This superficially abstract point can be readily
illustrated. Suppose a study intends to
take  measurements on $n$ patients $Y_1,\dots,Y_n,$ from
a population with known variance
$\sigma^2,$ and the estimator is the mean. If no data are missing,
then the statistical information about the mean is $n/\sigma^2.$
Now suppose that $n_d$ observations are missing. We will perform a
class-1 sensitivity analysis, so that the estimator is the mean for
both our primary and sensitivity analysis. Our primary analysis
will assume data are missing completely at random, and the sensitivity
analysis will assume that the missing values are from patients with
the same mean, but a different variance, $\sigma^2_m.$ 

Under our primary analysis assumption, we can obtain valid inference by
calculating the mean of the $n-n_d$ observed values, or by
using multiple imputation for the missing values. In both cases the 
information about the mean is the same:  $(n-n_d)/\sigma^2.$

Under our class-1 sensitivity analysis, we multiply impute the missing
data under our assumption, and again our estimator is the mean.
Now, however, the statistical information will be approximately $n^2/\{(n-n_d)\sigma^2 +
n_d\sigma^2_m\}.$ Further, the information about the mean from the sensitivity analysis depends
on $\sigma^2_m.$ Since $\sigma^2_m$ is
not estimable, this information is under the control of
the analyst.

\begin{figure}
\centering
\includegraphics[width=8cm]{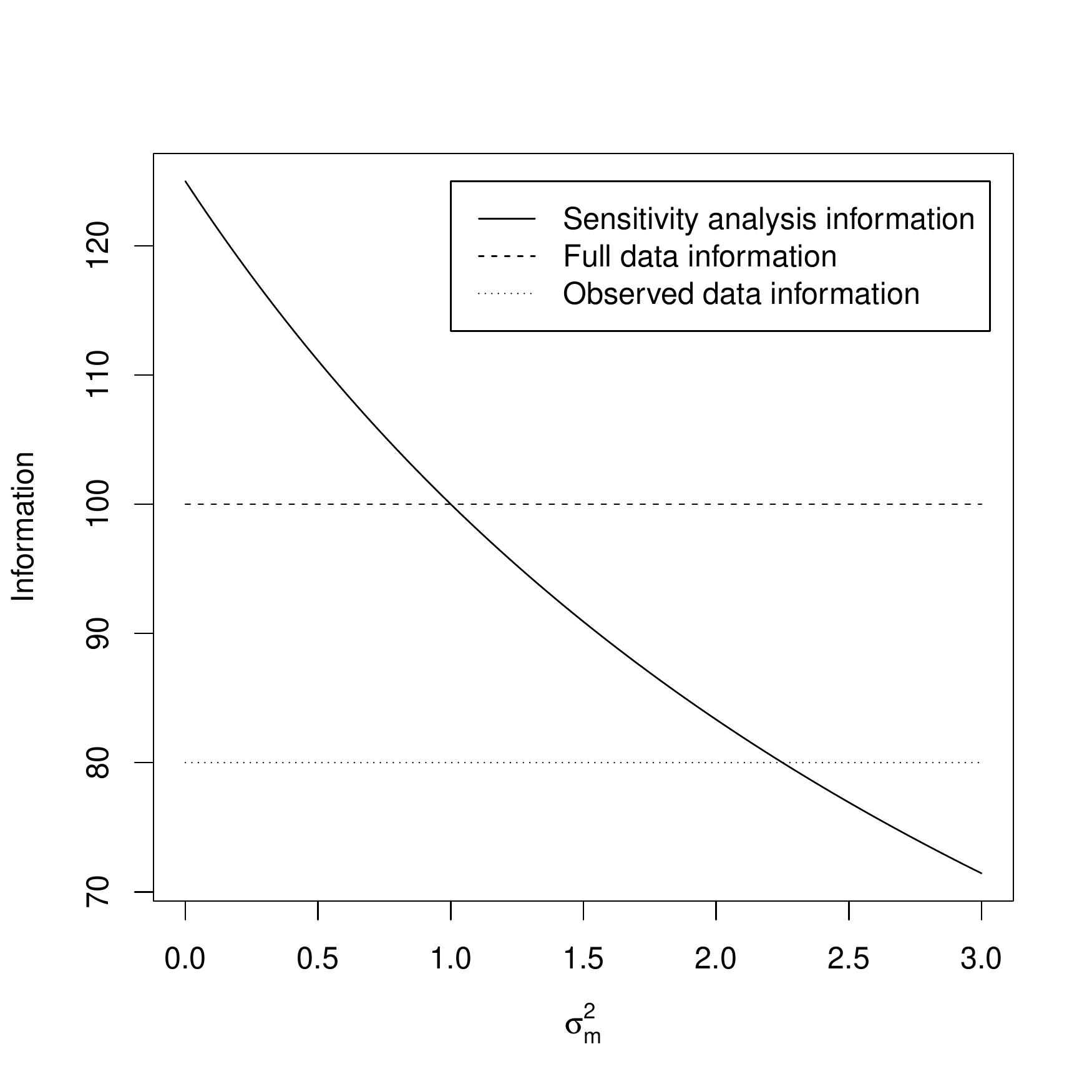}
\caption{\label{inf1}Information
about the sample mean varies with $\sigma^2_m.$}
\end{figure}

This is illustrated by Figure \ref{inf1}, which shows
how the information about the mean varies with
$\sigma^2_m,$ when $n=100,$ $n_d=20$ and $\sigma^2=1.$
When $\sigma^2_m< \sigma^2,$ the information about
the mean in the sensitivity analysis is greater than from the intended 100 observations;
when $1\leq\sigma^2_m\leq2.25$ then
the information is greater than in the $(n-n_d)$ observations
we were able to obtain, and
when $\sigma^2_m>2.25,$ the information is less than in the
observed data $(n-n_d)$ observations we were able to obtain.

We believe the ICH E9 (R1) addendum (\citeyear{ICH:2017})  will lead to
sensitivity analysis playing a much more central role;
in this context we believe it important for statisticians
and regulators to be aware of how---compared to
the primary analysis---information can be
removed or added in the sensitivity analysis.

Our purpose in this paper is to:
\begin{enumerate}
\item[1.] Consider the information in sensitivity analyses, arguing
  that sensitivity analysis in a clinical trial should
be information-anchored---as defined below---relative to the primary
analysis, and 
\item[2.] Demonstrate that using reference- and $\delta$-based controlled multiple
imputation, with Rubin's rules, to perform class-1 sensitivity
analyses is information-anchored.
\end{enumerate}

An important practical consequence of our work is that it provides a
set of conditions that can be imposed on class-1 sensitivity analyses
to ensure that---relative to the primary analysis---they neither
create, nor destroy, statistical information. We believe this provides
important reassurance for their use, for example in the regulatory setting. 

The plan for the rest of the paper is as follows. Section 2 defines
the concept of information-anchoring in sensitivity analysis.  Section
3 considers class-1 sensitivity analysis by reference- and $\delta$-based controlled multiple
imputation, and presents our main theoretical results on
information-anchoring within this setting. Section
4 briefly reviews class-2 sensitivity analyses from this perspective. In Section 5  we present a simulation study which illustrates our theory for information-anchored sensitivity analysis, which is then applied to a trial of training for peer reviewers in Section 6. We conclude with a discussion in Section 7.

\section{Information-Anchored Sensitivity Analysis \label{sec:simple_example}}

We have seen in the simple example above how a sensitivity analysis
can change the statistical information about a treatment estimate.
We now define {\em information-anchored} sensitivity analyses, which
hold the proportion of information lost due to missing data constant
across the primary and sensitivity analyses.

Suppose that a clinical trial intends to collect data from $2n$
patients, denoted $\bY,$ in order to estimate a treatment effect $\theta.$
However, a number of patients do not give complete data. Denote
the observed data by $\bY_{obs},$ and missing data by $\bY_{miss}.$ Consistent with the ICH-E9 (R1) addendum (\citeyear{ICH:2017}),
we make a {\em primary} set of assumptions, under which we perform the
primary analysis. We then make a {\em sensitivity} set of assumptions,
under which we perform the sensitivity analysis. Both primary and
sensitivity assumptions (i) specify the distribution
$[\bY_{miss}|\bY_{obs}],$ (ii) could be true, yet (iii) cannot 
be verified from $\bY_{obs}.$  

Let $\hat\theta_{obs,\,primary}$ be the estimate of $\theta$ under the
primary analysis assumption. Further, suppose we were able to
observe a realisation of $\bY_{miss}$ under the primary assumption. Putting these
data together with $\bY_{obs}$ gives us a complete set of observed
data, which actually follows the primary assumption: we denote this by
$\bY_{primary},$ and the corresponding estimate of $\theta$ by
$\hat\theta_{full,\,primary}.$ We denote the observed 
information about $\theta$ by $I(\hat\theta_{obs,\,primary})$ and $I(\hat\theta_{full,\,
  primary})$, respectively. Then,
\[
\frac{ I(\hat\theta_{obs,\,primary})}{I(\hat\theta_{full,\,primary})} < 1 ,
\]
reflecting the loss of information about $\theta$
due to
missing data.

Defining corresponding quantities under the sensitivity assumptions for the chosen sensitivity analysis procedure (be this class-1 or class-2) we
have,
\[
\frac{ I(\hat\theta_{obs,\,sensitivity})}{I(\hat\theta_{full,\,sensitivity})} < 1 ,
\]
again reflecting the loss of information about $\theta$
due to missing data---but now under the sensitivity assumptions.

Comparing these leads us to the following definitions,
\begin{eqnarray}
\notag
  \frac{I(\hat\theta_{obs,\,primary})}{I(\hat\theta_{full,\,primary})} > 
\frac{I(\hat\theta_{obs,\,sensitivity})}{I(\hat\theta_{full,\,sensitivity})}: 
&\mbox{\quad Information-{\em negative} sensitivity analysis,}  \\
\label{eq:iadef}
  \frac{I(\hat\theta_{obs,\,primary})}{I(\hat\theta_{full,\,primary})} = 
\frac{I(\hat\theta_{obs,\,sensitivity})}{I(\hat\theta_{full,\,sensitivity})}: 
& \mbox{\quad Information-{\em anchored} sensitivity analysis,}  \\
\notag
  \frac{I(\hat\theta_{obs,\,primary})}{I(\hat\theta_{full,\,primary})} < 
\frac{I(\hat\theta_{obs,\,sensitivity})}{I(\hat\theta_{full,\,sensitivity})}: 
& \mbox{\quad Information-{\em positive} sensitivity analysis.}
\end{eqnarray}

When analysing a clinical trial, we believe an information-positive sensitivity
analysis is rarely justifiable, implying as it does that the more data
are missing, the more certain we are about the treatment effect under
the sensitivity analysis. Conversely, while information-negative
sensitivity analyses provide an incentive for
minimising missing data, there is no natural consensus about the
appropriate loss of information. Therefore, we argue that information-anchored sensitivity analyses are the natural starting point. In
regulatory work they provide a level playing field between regulators
and industry, allowing the focus to be on the average response to
treatment among the unobserved patients.

The definitions above are quite general, applying directly to class-1
and class-2 sensitivity analyses, and all types of
{\em de jure} (on-treatment) and {\em de facto} (as-observed) assumptions. We now discuss 
class-1 sensitivity analyses from the information perspective and present our theory for information-anchoring.

\section{Class-1 Sensitivity Analysis and Theory for Information-Anchoring \label{sec:section_4}}

While class-1 sensitivity analyses can be performed without using
multiple imputation \cite[][]{Kaifeng, Lu:Pang, Tang:2017}, multiple
imputation is the most flexible approach, and often the simplest to
implement (e.g.,\ using the SAS software from
\underline{www.missingdata.org.uk} or Stata software by
\cite{Cro/Morris/Kenward/Carpenter:2016}). This is generally called
{\em controlled multiple imputation}, because the form of the imputation for
the missing data is {\em controlled} by the analyst. So, for example,
the analyst can control the imputed data mean to be $\delta$ below that
under missing at random (MAR). See, for example,  \cite{Mallinckrodt:2013} Ch.\ 10,
\cite{OKelly/Ratitch:2014}, p.~284--319   and \cite{Ayele/Lipkovich/Molenberghs/Mallinckrodt:2014}.

One approach is to obtain information about parameters that control the
departure from MAR from experts
\cite[][]{Mason/Gomes/Grieve/Ulug/Powell/Carpenter:2017}, but this is
controversial \cite[]{Heitjan:2017}, and challenging for longitudinal
data where multiple parameters are involved. An alternative,
  as introduced by \cite{Little/Yau:1996} and developed and discussed
  further more recently by, among others,
  \cite{Carpenter/Roger/Kenward:2013,Ratitch/OKelly/Tosiello:2013,Lu:Pang}, is {\em reference-based}
  multiple imputation. In this approach, the distribution of the
  missing data is specified {\em by reference} to other groups of
  patients. This enables contextually relevant qualitative assumptions to be explored and avoids the need to formally specify numerical sensitivity parameters (these are implicit consequences of the appropriate reference for a
patient). Some examples are listed in Table \ref{tab:options}.  For
  example, we may explore the consequences of patients in an active
  arm `jumping to reference' post-deviation. In practice the appropriate imputation model depends critically on the particular clinical setting and what assumptions are considered credible. Such analyses can be performed
  using the reference-based MI algorithm in Appendix \ref{app:alg}
  implemented in \cite{Cro/Morris/Kenward/Carpenter:2016}. 
Overall, this approach is  both very flexible, and accessible, since
  patients' missing outcomes are specified qualitatively---by
  reference to other groups of patients in the study. This explains
  its increasing popularity \cite[][]{jamapsychiatry20152146,JCPP:JCPP12443,Billingsdc171114,jama201720373}.

The above papers all focus on clinical trials with continuous outcome
measures that are collected longitudinally, and modelled using the
multivariate normal distribution.  We
consider the same setting, and give criteria for class-1 sensitivity analysis using controlled multiple
imputation with Rubin's variance formula to be information-anchored. 
This shows that most forms of $\delta$- and reference-based
imputation proposed in the literature are, to a good approximation,
information-anchored. It also shows that, in class-1 settings,
uncritical use of the conventional primary analysis variance estimator
is often information-positive, which is undesirable in practice. 

There are two principal reasons for this. The first is that
class-1 sensitivity analyses retain the primary analysis model in the
sensitivity analysis. However, in the sensitivity analysis data
assumptions are not wholly compatible with those of the primary
analysis model. In particular variance estimators may behave in
unexpected ways. The second reason is that reference-based methods
essentially use the data twice,  for example, by using data from the
reference arm (i) to impute missing data in an active arm and (ii) to
estimate the effect of treatment in the reference arm.

\subsection{Theoretical Results}
The presentation of our theoretical results is structured as follows. We begin by describing our data, model, primary
analysis and sensitivity analysis. We show in Corollary 2 that, when all data can be fully observed, for our
treatment estimate $\hat\theta,$ 
$$ \E[\vsfulhat] = \E[\vpfulhat] + O(n^{-2}).$$
Theorem 1 then defines the information-anchored variance and derives
a general expression for the difference between this and the variance
from Rubin's rules. Finally, we show, in the remarks following the
theorem,  that in practice this difference
is small.

\begin{table}
\caption{ \label{tab:options}Examples of reference-based
 and external information controlled multiple \mbox{imputation methods.}}
\fbox{%
\begin{tabular}{ L{4cm} | L{9cm} } \hline
Name & Description \\ \hline
Reference-based controlled MI methods: & \\
Jump to reference (J2R) & Imputes assuming that following dropout a
                          patients mean profile  follows that observed
                          in the   reference arm. Pre-drop  out means
                          come from the randomised arm. \\  
&\\
Copy increments in reference (CIR) & Forms post-dropout means by
                                     copying increments in the
                                     reference arm. Pre-drop out means
                                     come from the  randomised arm.\\
&\\
Last mean carried forward (LMCF) & Forms post-dropout means by
                                   carrying forward the randomised arm
                                   mean at dropout. \\
&\\
Copy reference (CR) & The conditional profile given the history is copied from the reference group  i.e. imputes as if randomised to reference   arm, pre- and post-drop out means come from the reference arm. \\
&\\
External information controlled MI methods: & \\
The $\delta$-method  & Impute under randomised arm MAR and
                       subtract/add by fixed $\delta.$\\ 
\end{tabular}
}
\vspace*{-0.1cm}
\end{table}

\subsection*{Trial Data}
Consider a two-arm trial, which includes $n$ patients randomised to an
active arm and $n$ patients randomised to a reference arm (total $2n$
patients within the trial). Outcome data are recorded at $j=1,...,J$
visits, where visit $j=1$ is baseline.   For patient $i$ in treatment
arm $z$, where $z=a$ indicates active arm assignment and  $z=r$
indicates reference arm assignment, let $Y_{z,i,j}$ denote the outcome
at time $j$.  

We wish to estimate the treatment effect at the end of the follow-up,
time $J.$ Our analysis model is the regression of the outcome at time
$J$ on treatment and baseline (i.e.,\ ANCOVA). Now suppose a number of patients
are lost to follow-up in the active arm (for simplicity, we assume for
now the reference arm data are complete). Our primary
assumption is MAR. 

Our primary analysis uses all the observed values,
imputes the missing data under MAR, fits the ANCOVA model to each
imputed data set and combines the results (this is essentially equivalent to fitting a mixed model with unstructured mean and covariance matrix to the observed values, see \citeauthor{Carpenter/Kenward:2008} (\citeyear{Carpenter/Kenward:2008}), Chapter 3).  

Our sensitivity analysis uses controlled multiple imputation, as formally
defined below. This could include a $\delta$-based method or
one of the reference-based methods given in Table \ref{tab:options};
all reference-based MI methods can be implemented using the
generic algorithm in Appendix \ref{app:alg}.

For each trial arm, we assume a multivariate normal model, with common
covariance matrix,
so that for patient $i$ who has no missing values: 
\[
\begin{pmatrix}
Y_{z,i,1}\\Y_{z, i,2}\\ \vdots \\ Y_{z,i,J}
\end{pmatrix}
\sim
N\left\{
\begin{pmatrix}
\mu_{z,1}\\\mu_{z,2}\\ \vdots \\ \mu_{z,J}
\end{pmatrix} ,
\bSigma = \left[
\begin{array}{cccc}
\sigma_{1,1}^2 & \sigma_{1,2}^2 & \hdots & \sigma_{1,J}^2 \\
 \sigma_{1,2}^2 & \sigma_{2,2}^2 & \hdots & \sigma_{2,J}^2 \\
\vdots & \vdots & \hdots & \vdots \\
 \sigma_{1,J}^2 & \sigma_{2,J}^2 & \hdots & \sigma_{J,J}^2 
\end{array}
\right]
\right\},
\]
where $z=a$ for the active patients and $z=r$ for the
reference patients. 

Now suppose all reference group patients follow the protocol, but 
$n_d=n-n_o$ active patients deviate from the protocol. Suppose it was
possible to continue to observe these
$n_d$ patients, 
but now their post-deviation data follows the controlled model:
 \begin{equation}
\label{eq:rbgen}
\begin{pmatrix}
Y_{a,i,1}\\ \vdots \\ Y_{a,i,j-1} \\ Y_{a,i,j} \\Y_{a,i,j+1}\\ \vdots \\ Y_{a,i,J}
\end{pmatrix}
\sim
N\left\{
\begin{pmatrix}
\mu_{a,1}\\ \vdots \\\mu_{a,j-1} \\ \mu_{d,j,j} \\
\mu_{d,j,j+1}\\
\vdots \\ \mu_{d,j,J}
\end{pmatrix} ,
\bSigma
\right\}.
\end{equation}
The term `controlled' means that the analyst controls the post-deviation
distribution. Here, for patient $i,$ the first index indicates active/deviation,
the second the time of deviation, and the third the visit number. 
Different patients can deviate at different times, and  this
general formulation allows the pattern of
their post-deviation means to differ depending on their deviation time. This
encompasses all the settings in Table \ref{tab:options}, and others besides.

To present the theory, we first consider the case where the primary
analysis does not adjust for baseline, extending to the baseline-adjusted case in Corollary 2. 

\medskip
\noindent{\sc Proposition 1}\\
For the trial data described above, when the analysis model is a
difference in means at the final time point with the usual sample
variance estimate in both observed and controlled settings, then:
\begin{itemize}
\item[(a)] If all patients follow the protocol and no data are
  missing, then the expectation of the variance estimate is:
\[
\E [\vpfulhat] = \frac{2\sigma_{J,J}^2}{n}.
\]
\item[(b)] 
If $n_d$ patients deviate and are observed following the controlled
model \eqref{eq:rbgen} the expectation of the variance estimate is:
\[
 \E [\vsfulhat] = \frac{2\sigma_{J,J}^2}{n} + \sum_{j=2}^{J}
\frac{n_on_{d,j}\Delta^2_{d,j}}{n^3}
+\sum_{p=2}^{J} \sum_{q=2}^{\overset{J}{q\neq p}} \frac{n_{d,p}n_{d,q}\Delta_{d,p,q}^2}{n^3},
\]
where $\Delta_{d,j}=\mu_{a,J}-\mu_{d,j,J},$
$\Delta_{d,p,q}=\mu_{d,p,J}-\mu_{d,q,J}$ and we let $(n-1)\rightarrow n$.
\end{itemize}

\noindent{\sc Proof:} Appendix \ref{app:proof1}.\hfill$\Box$

\medskip
\noindent{\sc Corollary 1}\\
For clinical trials designed to detect a difference 
of $\mu_{a,J}-\mu_{r,J}=\Delta,$ with a significance level of $\alpha$ and power
$\beta,$ at the final visit, $J,$
\[
\E [\vsfulhat] = \E [\vpfulhat]  + O(n^{-2}).
\]
{\sc Proof:}\\
First notice that the standard sample size formula implies
\[
\Delta^2 = \frac{2 f(\alpha,\beta) \sigma^2}{n}.
\]
Therefore, $\Delta^2$ is $O(n^{-1}).$ Further, since
in any trial, all $\Delta^2_{d,p,J}$ can be written
as $\Delta^2_{d,p,J} = \kappa_{d,p,J}\Delta^2$ for
some constant $\kappa_{d,p,j},$ we have $\Delta^2_{d,p,J} = O(n^{-1}).$ Following the same arguments,  $\Delta^2_{d,j} = O(n^{-1}).$
Second, notice that $n_o/n$ is the proportion of
active patients who complete the
trial, and $n_{d,j}/n$ is the proportion who deviate at time $j.$
Therefore, $n_on_{d,j}/{n^2} < 1.$ Similarly  $n_{d,p}n_{d,q}/{n^2} < 1.$
It therefore follows that,
\begin{equation}
\label{eq:within}\E[\vsfulhat] = \E[\vpfulhat] + O(n^{-2}).
\end{equation}

\medskip\noindent{\sc Corollary 2}\\
Under the conditions of
Corollary 1, if the
primary analysis model is a linear regression
of the outcome at the final time point, adjusted for baseline, then \eqref{eq:within}
still holds.

\medskip\noindent{\sc Proof:}\\
Replace the unconditional variance, $\sigma_{J,J}^2$, with the variance conditional on baseline, 
$\sigma_{J.1}^2=\sigma_{J,J}^2-\left(\sigma_{1,J}^2\right)^2/\sigma_{1,1}^2$,
in the proof of Proposition 1.\hfill$\Box$

\medskip
\noindent We now use this result in the context of reference-based
multiple imputation to calculate the difference between our defined
information-anchored variance and Rubin's multiple imputation variance.

\medskip
\noindent {\sc Theorem 1}\\
Consider a two-arm trial which includes $n$ patients randomised
to an active arm and $n$ patients randomised to a reference arm. Measurement data is recorded at $j=1,\dots, J$ visits (where visit 1 is
baseline).  The primary analysis model is a linear
regression of the outcome at the final
time point (visit $J$) on baseline outcome and treatment. Suppose all $n$ of the reference arm are completely observed on reference treatment over the full duration of the trial (at all $J$ visits)
but in the active arm, only $n_o$ are observed without deviation. The remaining $n_d$ patients in the active arm  deviate at some point during the trial post-baseline in a monotone fashion (such that $n_o+n_d=n$). Specifically, we assume a
proportion $\pi_{d,j}=n_{d,j}/n$ drop out at each visit, for $j>1$ and their data are missing post-deviation. 

Assume that the primary design-based analysis model satisfies \eqref{eq:within},
and that the variance covariance matrix for the data is the same in each
arm. For the patient deviation pattern in the active arm  beginning at time $j$, 
let $\bar{\bP}_{a,d,j}$ be the $j\times 1$ mean vector of the
  $n_{d,j}$ responses at times $1,\dots,(j-1)$ plus a 1 (to allow for
  an intercept in the imputation model). 

Suppose the primary analysis is performed
by MI assuming
 within-arm MAR. Let $\vpmshat$ denote the estimated variance for the treatment effect under the primary MAR assumption. Subsequently 
we perform class-1 sensitivity analysis via reference-based MI, i.e.\
under \eqref{eq:rbgen}, using the
imputation algorithm in Appendix A. This general formulation
includes all the reference-based options in Table \ref{tab:options}.  As we are doing class-1
sensitivity
analysis, the primary analysis model is used to analyse the imputed
data. Then the difference between the
information-anchored variance of the sensitivity analysis
treatment estimate, denoted by $\vanchat$, which by definition is 
$(\vpmshat / \vpfulhat) \times \vsfulhat$ and Rubin's MI variance, denoted by $\vrubin$, is
\begin{equation}
\label{eq:ia} \bE[\vanchat]-\bE[\vrubin] =  \sum_{j=2}^{J}
\pi^2_{d,j}\bar\bP_{a,d,j}(\bV_{primary,j} - \bV_{sensitivity,j})
\bar\bP^T_{a,d,j}
+ \frac{\bE[\bprimary]}{O(n^2)\bE[\wprimary]}.
\end{equation} 

Here $\bV_{primary,j}$ is the variance-covariance matrix
of the parameter estimates in the primary MAR imputation model
for deviation at time $j$ 
and $\bV_{sensitivity,j}$ is the variance-covariance matrix of
 the  parameter estimates in the imputation model for deviation at time $j,$  defined by the reference-based sensitivity analysis assumption. $\bprimary$ is the between-imputation variance and $\wprimary$ is the within-imputation variance of the treatment effect in the primary analysis, both under MAR.

\medskip
\noindent {\sc Proof:} Appendix \ref{app:proof2}\hfill$\Box$

\medskip
\noindent Theorem 1 establishes the difference between the
information-anchored variance and Rubin's rules variance.
To show that class-1 sensitivity analysis
by reference-based multiple imputation is information-anchored, we need to consider how close expression \eqref{eq:ia} is to zero.

The key quantity driving the approximation is the first of the two terms.
Notice that for each deviation time, $j,$
the variance covariance matrix of the parameters of the
on-treatment imputation model  is
 $\bV_{primary,j}=\bSigma_j/n_o,$
where $\bSigma_j$ is the relevant
sub-matrix of the variance-covariance matrix $\bSigma$ of the $J$ 
observations. The precise form of $\bV_{sensitivity,j}$ will depend
on the sensitivity analysis imputation model. Consider data from the fully observed reference arm are used
in the sensitivity imputation (e.g.\ copy reference). In this case, 
 $\bV_{sensitivity,j}=\bSigma_j/n,$ and
\begin{eqnarray}
\pi^2_{d,j} \bar\bP_{a,d,j} [\bV_{primary,j} - \bV_{sensitivity,j}]\bar\bP^T_{a,d,j} &=
\pi^2_{d,j} \bar\bP_{a,d,j}\bSigma_j
\left[\frac{1}{n_{o}}-\frac{1}{n}\right]\bar\bP^T_{a,d,j} \nonumber\\
&= \pi_{d,j}^2\bar\bP_{a,d,j}\bSigma_j\left[\frac{n-n_{o}}{n_{o} n}\right]\bar\bP^T_{a,d,j} \nonumber\\
&=\pi_{d,j}^2\bar\bP_{a,d,j}\bSigma_j\left[\frac{\pi_{d}}{n\left(1-\pi_{d}\right)}\right]\bar\bP^T_{a,d,j}. \nonumber
\end{eqnarray}
Applying this line of argument to the
other methods in Table \ref{tab:options}
 suggests that the error in the approximation will be small,
and vanish asymptotically. 

Thus we have established that
class-1 referenced-based imputation sensitivity analysis is,
to a good approximation, information-anchored. We illustrate this
in the simulation study in Section \ref{sim_study}.

\subsection{Further Comments}
\begin{enumerate}
\item In the proof of
Theorem 1, to simplify the argument,  the variance-covariance matrix of the
data $\bSigma$  is assumed known {\em in the
imputation model.} When---as will generally be the case---it has to be estimated, \citeauthor{Carpenter/Kenward:2013}
(\citeyear{Carpenter/Kenward:2013}), p.~58--59, show that, for
the simple case of the sample mean, the additional
bias is small, and vanishes asymptotically. This strongly suggests
that any additional bias caused by estimating the variance
covariance matrix will be small, and asymptotically irrelevant; this
is borne out by our simulation studies below.
\item[]
\item For simplicity the theory  treated the deviation pattern as fixed. We can replace all the proportions, $\pi_{d,j}$ by their sample
estimates, and then take expectations over these in a 
further stage. As our results are asymptotic, the conclusions will be
asymptotically equivalent. 
\item[]
\item $\delta$-method sensitivity analysis: We consider that at the final time point $J$ imputed values for patients who deviate at time $j$ (for $j>1$) are edited by $(J+1-j)\delta$ to represent a change in the rate of response of $\delta$ per time point post-deviation. We now evaluate the size of the two terms in \eqref{eq:ia} separately. For the first term, when $\delta$ is fixed, the covariance matrix
for the imputation coefficients under the primary analysis and the
sensitivity analysis is identical for each missing data pattern $j$; the $\delta$-method simply adds
a constant to the imputed values. Consequently
$\bV_{primary,j}=\bV_{sensitivity,j},$ thus $ \pi^2_{d,j}\bar\bP_{a,d,j} [\bV_{primary,j} -
\bV_{sensitivity,j}]\bar\bP^T_{a,d,j}=0,$ and 
Rubin's rules give a very sharp approximation to the
information-anchored variance. \\

However when $\delta$ is not fixed and we vary $\delta$ over the imputation set $K$, that is we suppose $\delta_k\sim N\left(\delta, \sigma_{\delta}^2\right)$, then, $ \pi^2_{d,j}\bar\bP_{a,d,j} [\bV_{primary,j} -
\bV_{sensitivity,j}]\bar\bP^T_{a,d,j}=-\pi_d^2\sigma_{\delta}^2$, and
the sensitivity analysis is {\em information-negative}. The extent
of this is principally driven by the variance of $\delta_k$. \\

Now consider the second term in \eqref{eq:ia}. When the $\delta$-method is used it is not necessarily the case that  \eqref{eq:within} holds, since $\Delta_{d,j}=\mu_{a,J}-\mu_{d,j,J}$ and
$\Delta_{d,p,q}=\mu_{d,p,J}-\mu_{d,q,J}$ are not necessarily  $O(n^{-1})$. In the $\delta$-based scenario, as outlined in Appendix \ref{app:proof1},
\begin{eqnarray}
\vsfulhat = \vpfulhat + Q,\nonumber
\end{eqnarray}
where,
\begin{eqnarray}
Q = \sum_{j=2}^{J}\frac{n_on_{d,j}\left(J+1-j\right)^2\delta^2}{n^3} + \sum_{p=2}^{J}\sum_{q=2}^{\left(J, q\neq p\right)}\frac{n_{d,p}n_{d,q}\left(\left(J+1-p\right)\delta-\left(J+1-q\right)\delta\right)^2}{n^3}.\nonumber
\end{eqnarray}

Thus, for the $\delta$-method the $O(n^{-2})$ component in the second term of \eqref{eq:ia} is replaced with Q (as defined above). The composition of Q indicates that the information-anchoring performance of Rubin's variance estimate will also depend on the size of $\delta$. Typically, the size of $\delta$ will not have a large effect since the terms in Q are all multiplied by components of the form $n_on_{d,j}/n^3$ or $n_{d,p}n_{d,q}/n^3$ and thus will vanish asymptotically. Hence with a fixed $\delta$ adjustment, the information-anchoring approximation will be excellent.

\item[]

\item Improved information-anchoring: Remark  (b) shows that, provided the
underlying variance-covariance matrices of the data are 
similar, the key error
term in the information-anchoring approximation is
the difference in precision with which they are estimated.
If all $n$ patients are observed in the reference arm and $n_o$ in the active arm, this is 
\[
\frac{1}{n_{o}} - \frac{1}{n}.
\]
This suggests Rubin's rules will lead to
improved information-anchoring if, instead of
using all patients in the reference arm to
estimate the imputation model for deviators at time $j,$
a random $n_{o}$ are used. We have confirmed this
by simulation, but the improvement is negligible when the
proportion of missing data is $<40\%$, when simulations
confirm the approximation is typically excellent.
\item[]
\item Theorem 1 suggests that,
for a given deviation pattern, information-anchoring will
be worse the greater the difference between the
covariance matrix of the imputation coefficients under the primary 
and sensitivity analysis. However, we have not encountered examples where
this has been a practical concern.
\item[]
\item We have not presented formal extensions of our theory to the
case when we also have missing data in the reference arm. But this does
not introduce any substantial errors in the information-anchoring approximation.
With missing data in the reference arm, for each missing data pattern $j$, an additional component which depends on the difference between the variance of the imputation
parameters in the primary on-treatment imputation model and sensitivity scenario imputation model for the reference arm,
multiplied by the proportion of reference patients with that missing data pattern squared (denoted $\pi_{r,d,j}^2$) is included.  If reference arm data are imputed under within-arm MAR (as under CIR, CR or J2R) these terms will  be zero. In the more general case, where different patterns of patients, across different arms,
are imputed with different reference-based assumptions, additional non-zero error terms of the form as in the summation in \eqref{eq:ia} will be introduced; but again, for the reasons discussed above, these will typically be small. The covariance between the parameters of the active and reference arm sensitivity scenario imputation models for each missing data pattern also contributes to the sharpness of the approximation. The exact size of these additional error terms again depend on the specific sensitivity scenario and in some cases will be zero (e.g.\ LMCF). But each covariance term is always multiplied by the proportion of deviators in each arm with the associated missing data patterns ($\pi_{d,j}\pi_{r,d,j}$), $\bar\bP_{a,d,j}$ and $\bar\bP^T_{r,d,j}$ (the $j\times 1$ mean vector of the responses at times $1,\dots,(j-1)$ for the reference patients deviating at time $j$, plus a 1 to allow for an intercept in the imputation model). Thus will be of a relatively small order in practice following the reasons discussed above.

\end{enumerate}

\subsection*{Summary}
Given a primary design-based analysis model,
we have established in Proposition 1 a criterion which
defines a general class of reference-based sensitivity analyses.
If these sensitivity analyses are performed by MI,
we have further established in Theorem 1 that they will
be---to a good approximation---information-anchored, in line
with the principles we set out in Section \ref{sec:simple_example}. We
have also shown why the information-anchoring is  particularly sharp for the $\delta$-method of MI.

\section{Class-2 Sensitivity Analyses and Information-Anchoring}

A full exploration of information-anchoring for class-2 sensitivity
analyses is beyond the scope of this article. Here, we focus on
likelihood-based selection models \cite[see,
for example,][]{Diggle/Kenward:1994}, and use the
results of \cite{Molenberghs/Michiels/Kenward/Diggle:1998} to make
links to pattern mixture models, which allows us to use the results 
we presented in Section \ref{sec:section_4}.

Continuing with the setting in Section \ref{sec:section_4},  consider a trial with
scheduled measurement times of a continuous outcome measure at
baseline and over the
course of the follow-up. When data are complete, the primary analysis
is the ANCOVA of the outcome measure at the scheduled end of follow-up on
baseline and treatment group. Equivalent estimates and inferences can
be obtained from a mixed model fitted to all the observed data,
provided we have a common unstructured covariance matrix and a full
treatment-time and baseline-time interaction. 

Now suppose patients withdraw before the scheduled end of follow-up,
and subsequent data are missing. The mixed model described in the
previous paragraph then provides valid inference under the assumption
that post-withdrawal data are MAR given baseline, treatment group and available
follow-up data. A selection model that allows post-withdrawal data to be MNAR
combines this mixed model with a model for the dropout process. Let $R_{i,j}=1/0$ if
we observe/miss the outcome for patient $i$ at scheduled visit
$j=1,\dots,J.$ An illustrative selection model is:
\begin{eqnarray}
\notag
Y_{i,j} & = \alpha_j + \beta_j Y_{i,0} + \gamma_jT_{i} + \epsilon_{i,j},
\quad \beps_{i}\sim N(\bzero, \bSigma_{{\textnormal{\tiny
          J}}\times{\textnormal{\tiny J}}} ) \\
\label{eq:sel} g(R_{i,j}) & = \alpha^R_j + \beta^R_j Y_{i,0} +
                           \gamma^R_jT_{i} + \delta^R_1 Y_{i,j-1} + \delta^R_2(Y_{i,j}-Y_{i,j-1}),
\end{eqnarray}
where the superscript `$R$' denotes a selection model parameter, and the link function $g$ is typically logit, probit or
complementary log-log (the latter giving a discrete time proportional hazards
model for withdrawal).

Usually there is little information on the informative missingness
parameter
 $\delta^R_2$ in the data
\cite[]{Rotnitzky/Cox/Bottai/Robins:2000,Kenward:1998}, and this
information will be highly dependent on the assumed data distribution. Therefore,
in applications it is  more useful to  explore the robustness
of inferences to specific, fixed, values of $\delta^R_2$ ($\delta^R_2=0$
corresponds to MAR).

For each of these specific values of $\delta^R_2,$ we may recast the
selection model as a pattern mixture model, following
\cite{Molenberghs/Michiels/Kenward/Diggle:1998}. The 
differences between the observed and unobserved patterns are defined
as functions of the fixed $\delta^R_2.$ However, these then become a
particular example of the $\delta$-method pattern mixture models considered in Section \ref{sec:section_4},
which we have shown are information-anchoring. 

More generally, local departures from MAR are asymptotically information-anchored. To see this, denote by $\bthe$ the parameters in
\eqref{eq:sel}, apart from $\delta_2^R.$ For a fixed $\delta_2^R,$ let $\boldi(\btheh;\delta_2^R)$ be the
observed information matrix at the corresponding maximum likelihood
estimates $\btheh.$ For regular log-likelihoods and a given data set, as we move away from
MAR, for each element, $i,$ of the information matrix $\boldi,$ the
mean value theorem gives
\begin{eqnarray}
\label{eq:eam} i(\btheh;\delta_2^R) -  i(\btheh;0) =
\left(\left.\frac{\partial}{\partial\delta_2^R} i(\btheh;\delta_2^R)\right|_{\delta_2^R=\tilde\delta_2^R}\right){\delta}_2^R , \mbox{ for some } \tilde\delta_2^R \in (0, \delta_2^R).
\end{eqnarray}
However, asymptotically the parameter estimates are normally
distributed, so the third derivative of the likelihood (i.e. the RHS
of \eqref{eq:eam}) goes to zero. Because the above holds when we use
both the full data, and the partially observed data, it is sufficient to give
information-anchoring. This is the basis for our intuition that, for most Phase III trials, class-2
sensitivity analyses can be treated as information-anchored for practical purposes. 

\section{Simulation Study \label{sim_study}} 

We now present a simulation study which illustrates the information-anchoring property of Rubin's variance formula, derived in Section
\ref{sec:section_4}.  
The simulation study is based on a double-blind chronic asthma
randomised controlled trial conducted by
\cite{Busse/Chervinsky/Condemi/Lumry/Petty/Rennard/Townley:1998}. The
trial compared four doses of the active treatment budesonide against
placebo on forced expiratory volume (FEV$_1$ recorded in litres) over
a period of 12 weeks. FEV$_1$ measurements were recorded at baseline
and after 2, 4, 8 and 12 weeks of treatment. The trial was designed to
have 80\% power (5\% type-1 error) to detect a change of
0.23 litres in \fev\ with 75 patients per arm, assuming a SD of 0.5 litres.

We simulated longitudinal data, consisting of baseline and two
follow-up time points (time 2 being week 4, and time 3 being week 12),
from a multivariate normal distribution whose
mean and covariance matrix were similar to those observed in the
placebo and lowest active dose arm of this trial:  
\[
\begin{array}{c} \mathbf{\Sigma_{placebo}} = \mathbf{\Sigma_{active}} \end{array}=\left[\begin{array}{ccc}
0.4 & 0.2 & 0.2  \\
0.2 & 0.5 & 0.2 \\
0.2 & 0.2 & 0.6
\end{array}\right],
\]
\[
\mathbf{\mu_{placebo}}=\left[2.0, 1.95, 1.9\right],\quad
\mathbf{\mu_{active}}=\left[2.0, 2.21, \mu_{a, 3}\right] \quad \mbox{(litres).}
\]

In the asthma study $\mu_{a, 3} \approx 2.2$ litres, corresponding to
a treatment effect of $\approx 0.3$ litres at time 3 (week 12). In the
simulation study we explored  $\mu_{a, 3}=\left\{1.9, 2.2,
  2.9\right\}.$ To test the approximation \eqref{eq:within} we 
chose a sample size  of $n=250$ in each arm, giving a power of at least 90\% in all scenarios. 
  For each
scenario, the analysis model was a linear regression of \fev\ at visit 2 and
baseline and treatment, and this was fitted to the full data.

Subsequently, for the active arm, we
simulated monotone deviation completely at random. We varied the proportion of patients
deviating overall from 0-50\%. For each overall proportion deviating, around half the patients
deviated completely at random before visit 2, and around half deviated
completely at random before visit 3.  All
post-deviation data were set to missing. The reference arm was always fully observed. 

For each simulated data set, the primary analysis assumed MAR, and we
performed class-1 sensitivity analyses using each of the reference-based methods in Table \ref{tab:options}. Fifty imputations were used for each
analysis. For the $\delta$-method, the unobserved data was postulated to be worse (than under MAR)
by a fixed amount of $\delta=\left\{0, -0.1, -0.5, -1\right\}$, for
each time point post-deviation, where $\delta=0$ is equivalent to the
primary, MAR analysis. Thus, for patients who deviated between visits
1 and 2,
their MAR imputed observations at visit 2 were altered by $\delta$ and
at time 3 by $2\delta$. For patients who deviated between visits 2 and
3, their MAR imputed observation at time 3 was altered by $\delta$.

One thousand independent replicates were generated for each
combination of $\mu_{a, 3}$ and deviation. Our results focus on the
visit 3 treatment effect and its variance.

In order to minimise the Monte-Carlo
variability in our comparisons,
we used the same set of 1000 datasets and deviation patterns for
each sensitivity analysis. 

Within each replication, for each sensitivity scenario, we also drew post-deviation data under
this scenario, giving a complete scenario-specific data set. For each
replication this allowed us to
estimate the  treatment effect and  $\vsfulhat$ for each scenario. Then,
we calculated the theoretical information-anchored variance, which by
definition in Section \ref{sec:simple_example} is $\hat{V}_{\textnormal{\tiny anchored}}=(
\vpmshat / \vpfulhat)\times\vsfulhat$. Rubin's variance estimate was
calculated. Estimates were averaged over the 1000 simulations. 
All simulations were performed using Stata version 14 (\citeauthor{stata14}, \citeyear{stata14}) and reference-based MI was conducted using the mimix program by \cite{Cro/Morris/Kenward/Carpenter:2016}.

\subsection{Simulation  Results}

\begin{figure}
\centering
\makebox{\includegraphics{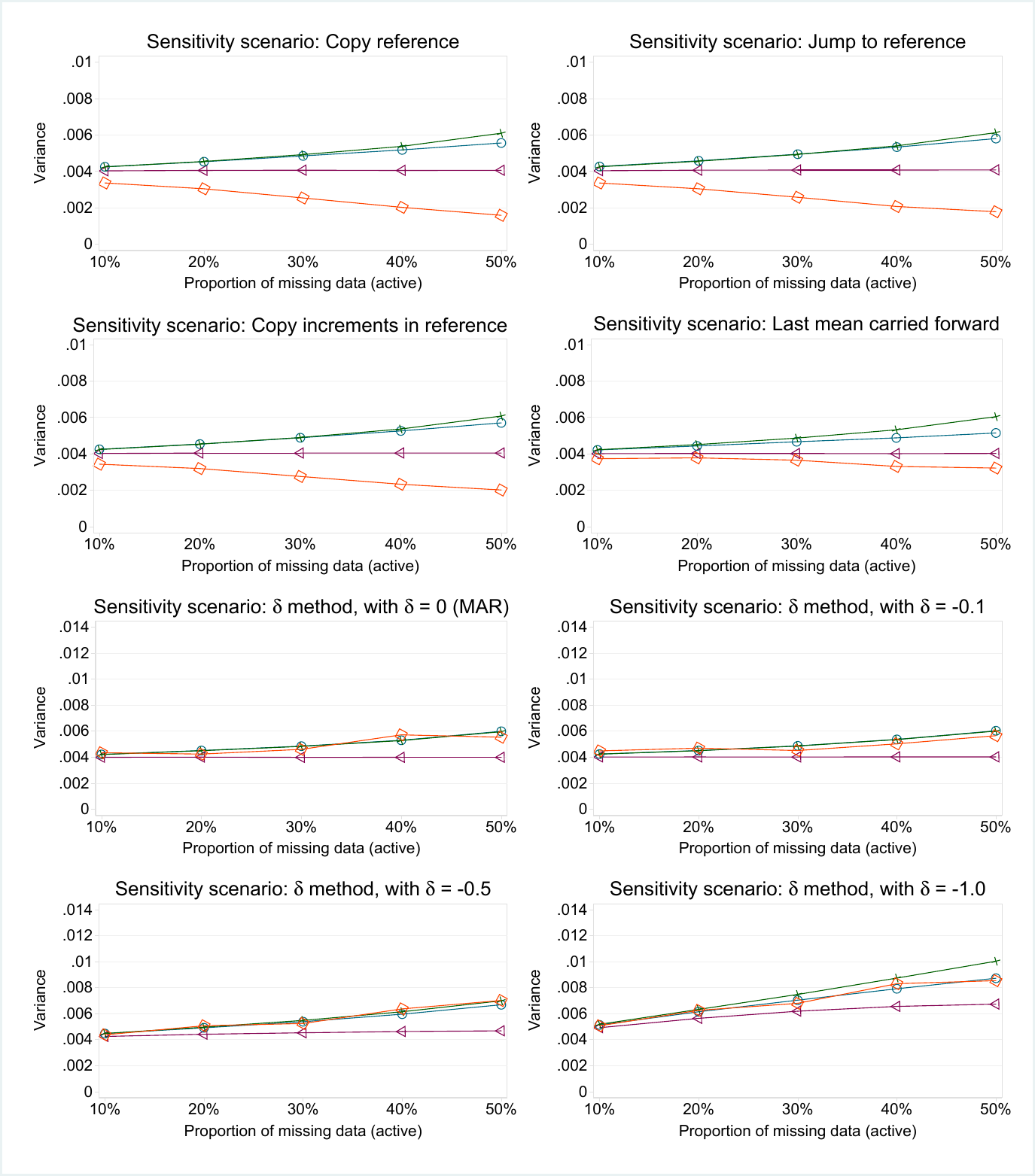}} 
\caption{\label{fig:simres}Simulation results: for each sensitivity
  scenario, as the proportion of
active arm deviations increases, each panel shows the evolution of the
mean estimate of the visit 3 treatment effect variance (over 1000 replications) calculated in
four ways:  (i) $-\!\!\circ\!\!-$ Rubin's MI variance, from
reference-/$\delta$-based MI;  (ii)  $-\!\!\times\!\!-$
information-anchored variance ($\hat\E[\hat{V}_{\textnormal{\tiny anchored}}]$);  (iii) $-\!\!\diamond\!\!-$ applying primary analysis
variance estimator in sensitivity scenarios; (iv) $-\!\!\triangleleft\!\!-$ Variance when post-deviation data is actually fully observed under the given scenario ($\hat\E[\vsfulhat]$).}
\end{figure}

Figure \ref{fig:simres} 
shows the results, for each of the reference-based sensitivity
scenarios in Table \ref{tab:options}, and controlled multiple imputation
with four values of $\delta.$ 

The top four panels are for a moderate
treatment effect of 
$0.3$ ($\mu_{a, 3}=2.2$),
comparable to that found
in the asthma trial. We see 
the results show excellent information-anchoring by Rubin's variance estimator for up to $40\%$
of patients deviating. Notice the information-anchored variance is
always greater than $\vsfulhat,$ the variance we would see if we were able to
observe data under the sensitivity assumption.

These results are echoed by those with  smaller and larger treatment
effects (see Appendix \ref{app:sims} Figure \ref{fig:simres2}). We conclude that, for realistic proportions of
missing post-deviation data, reference-based multiple imputation using Rubin's
variance estimator can be regarded as information-anchored.

This is in contrast to the behaviour of the conventional variance
estimator from the primary regression analysis. Across all four reference-based
scenarios, this gets smaller---and tends to zero---as the proportion of missing data
increases, so yields increasingly information-positive inference as more data are
missing! It is also smaller than the
variance we would obtain {\em if we were able to observe data under
the sensitivity assumption}.  Therefore, 
 (see \cite{Carpenter/Roger/Cro/Kenward:2014}) we believe this is not generally
an appropriate variance estimator for class-1  sensitivity 
analyses. We return to this point below.

Now consider the lower four panels of Figure \ref{fig:simres}, which show
results for controlled multiple imputation using the $\delta$-method.
Again,  consistent with the theory in Section \ref{sec:section_4}, these
show excellent information-anchoring by Rubin's variance estimator for
all missingness scenarios for $\delta=0, -0.1, -0.5$ litres. 
Indeed, the information-anchoring approximation is better than for the
reference-based methods above because the covariance matrix for the
imputation coefficients under MAR and $\delta$-based imputation are
identical: term 1 in \eqref{eq:ia} disappears. 

For contextually large $\delta=-1$ litres, the approximation is excellent for
up to 40\% missing data.  For greater proportions of missingness the
approximation is not so sharp, and this is caused by the size of the
second term in \eqref{eq:ia}, which is larger with a bigger $\delta$
and greater proportion of missing post-deviation data.  

For the $\delta$-method we also see using the conventional variance estimator from the
primary analysis is also information-anchored. The reason for
different behaviour here than for reference-based methods is that
reference-based methods borrow information from another trial arm, and they do this increasingly as the
proportion of patients deviating increases. This causes the
conventional variance estimator to be information-positive. However,
with the  $\delta$-method there is no borrowing between arms, so this issue does
not arise.

To summarise, the simulations demonstrate our theoretical results,
showing that for all the controlled MI methods outlined in Table
\ref{tab:options} (reference- and $\delta$-based),  in realistic trial
settings multiple imputation
using Rubin's rules gives information-anchored inference for treatment
effects. It is only with very high proportions of missing data
(e.g. $>50$\%) that the information-anchoring
performance of Rubin's variance begins to deteriorate.  
Such high proportions of missing data are unlikely in well designed trials, and would typically be indicative of other major problems.

\section{Analysis of a Peer Review Trial \label{sec:application}}

We now illustrate how the information-anchored theory outlined in
Section \ref{sec:section_4} performs in practice, using data from a
single blind randomised controlled trial of training methods for peer
reviewers of the {\it British Medical Journal}. Full details of the trial are given in  \cite{Schroter:2004}.

\subsection{Description of the Data}
Following concerns about the quality of peer review, the
 original trial was set up to evaluate no-training, face-to-face training or a self-taught training
 package.  After consent, but before randomisation, each participant
 was sent a baseline paper to review (paper 1) and the review quality
 was measured using the Review Quality Index (RQI). This is a
 validated instrument which contains eight items and is scored from 1
 to 5, where a perfect review would score 5. All 609 participants who
 returned their review of paper 1, were randomised to
 receive one of the three interventions. 

Two to three months later, participants  were sent a further article
to review (paper 2). If this paper was reviewed a third paper was sent
three months later (paper 3). Unfortunately,  not all of the
reviewers completed the required reviews, thus a number of review
scores were missing. The main trial analysis was conducted under the
MAR assumption, using a linear regression of RQI  on intervention
group adjusted for baseline RQI. The analysis showed that the only
statistically significant difference was in the quality of the review
of paper 2, where the self-taught group did significantly better than
the no-training group.

\begin{figure}
\centering
\makebox{\includegraphics[scale=1]{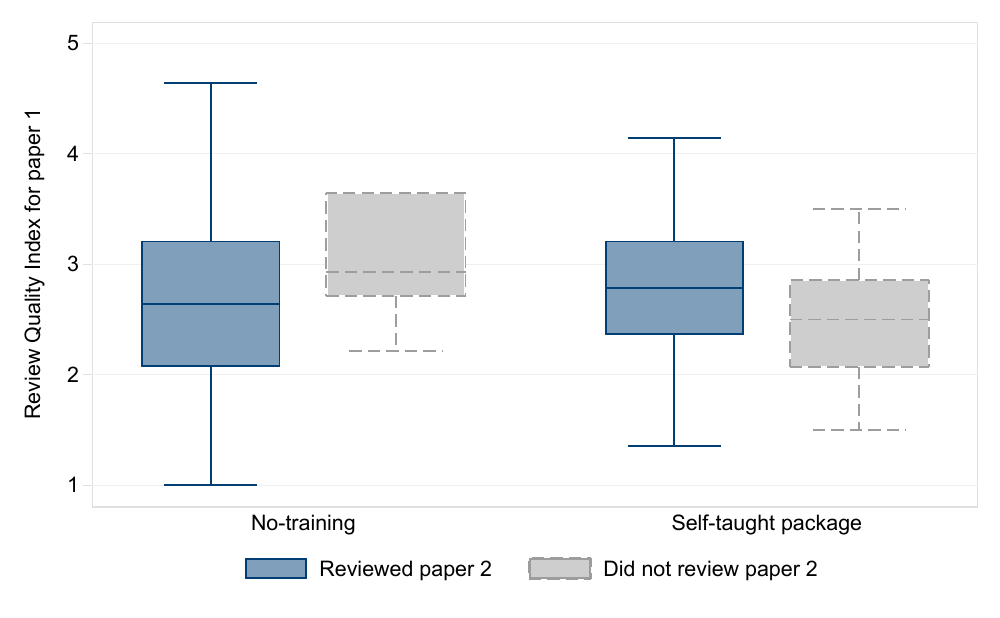}} \caption[The quality of the baseline review]{\label{fig_rev} The quality of the baseline review}
\end{figure}

Therefore, here we focus on examining the robustness of this purportedly significant result to
different assumptions about the missing data. Assuming MAR, the
analysis found that reviewers in the self-taught group  had
a mean RQI 0.237 points above the no-intervention group (95\% CI
0.01--0.37, $p=0.001$). Although this is relatively small, the
self-taught intervention is inexpensive and may be worth pursuing.
However,  Figure \ref{fig_rev} shows the quality of the review at
baseline for (a) those who went on to complete the second review and
(b) those who did not, for each of these two trial arms. The results
suggest that a disproportionate number of poor reviewers in the
self-taught group failed to review paper 2. This suggests the MAR
assumption may be inappropriate, and data may be missing not at random.

\subsection{Statistical Analysis}
The primary analysis model was a linear regression of paper 2 RQI on
baseline and intervention group (self-taught vs no-training), and the intervention effect estimate
is shown in the first row of Table \ref{tab:reviewer}.

We conducted four further analyses:
\begin{enumerate}
\item We multiply imputed the missing RQI data assuming MAR, fitted
  the primary analysis model to each imputed dataset and combined the
  results for inference using Rubin's rules. The imputation model for
  RQI of paper 2 included the variables present in the primary
  analysis model (RQI at baseline and treatment group).
\item As it is reasonable to suppose that many of the reviewers in the
  self-taught group who did not return their second review ignored
  their training materials, we perform a class-1 sensitivity analysis 
assuming they `copied no-training'. We used MI and Rubin's rules for
information-anchored inference.
\item We reproduced a previous sensitivity analysis described by
  \cite{White/Carpenter/Evans/Schroter:2007}.
They used a questionnaire  to elicit experts'
 prior opinion about the
average difference in review quality index between those who did, and
did not, return the review of paper 2 (20 editors and other staff at the BMJ
completed the questionnaire). The resulting distribution can be
summarised as 
$N(-0.21, 0.46^2)$. We used this to perform a $\delta$-method 
sensitivity analysis, where, for each imputation $k,$ RQI values in the self-taught arm
were imputed under  MAR  and then had $\delta_k\sim N(-0.21, 0.46^2)$
added. This analysis is expected to be information-negative.
\item Our fourth analysis used the $\delta$-method via MI for participants in
  the self-taught arm, but now fixed $\delta=-0.21$ (the mean expert opinion) to obtain information-anchored analysis.
\end{enumerate}
 All analyses used 50 imputations and were performed using Stata version 14 (\citeauthor{stata14}, \citeyear{stata14}).

\subsection{Results}
Table \ref{tab:reviewer} shows the results. As theory predicts, rows 1
and 2 show that the
primary analysis and analysis assuming MAR using MI give virtually
identical results.
In row 3, reference-based sensitivity analysis assuming copy no-training reduces
the estimated effect to 0.172; compared to the primary analysis the
information-anchored standard error (SE) is now very slightly reduced at 0.069. The
effect of this is to increase the p-value by a factor of ten to 0.013.

In contrast, using the expert's prior distribution (row 4), the point estimate
is 0.195, but the standard error is much increased at 0.132, so the
p-value is over 100 times greater than in the primary analysis.  
Lastly (row 5), again using the $\delta$-method, but
now fixing $\delta=-0.21$ gives a similar point estimate, but an
information-anchored SE of 0.072. 

Critically, comparing rows 4 and 5 shows that expert opinion 
loses a further 
$$\frac{\frac{1}{0.072^2}-\frac{1}{0.132^2}}{\frac{1}{0.072^2}} =70\% $$ 
of the information {\em beyond that lost due to missing data under the primary
analysis}. Such information losses are not atypical 
\cite[]{Mason/Gomes/Grieve/Ulug/Powell/Carpenter:2017}. Since trials are often powered with minimal regard to
potential missing data, such a loss of information must frequently
lead to the primary analysis being overturned. By contrast,
information-anchored sensitivity analysis fixes the loss of
information across the primary and sensitivity analysis, at a level
that is possible to estimate a-priori for any given deviation pattern.

\begin{table}
\caption{\label{tab:reviewer} Estimated effect of self-training vs no
  training on the paper 2 Review Quality Index, from the primary and
  various sensitivity analyses; $\dagger$ indicates information-anchored analysis.}
\fbox{%
\begin{tabular}{llll} \hline
Analysis & Est & SE & p-value \\ \hline
Primary analysis, MAR$^\dagger$ &  0.237 & 0.070 & 0.001 \\
\\
MI, MAR$^\dagger$ & 0.234 & 0.071 & 0.001 \\
\\
MI, copy no-training$^\dagger$ & 0.172  & 0.069 &  0.013 \\
\\
MI, expert opinion & 0.195 & 0.132 & 0.145 \\
$\delta_k \sim N(-0.21, 0.46^2)$ & & & \\
\\
MI, $\delta$-method with &  0.189 & 0.072 & 0.009 \\
$\delta=-0.21$$^\dagger$ &        &   &   \\
\hline
\end{tabular}
}
\end{table}

\section{Discussion \label{sec:discussion}}
The recent  publication of the ICH E9 (R1) addendum
(\citeyear{ICH:2017}) is  bringing a sharper focus on the
estimand. As the addendum recognises, this in turn leads to
greater focus on the assumptions underpinning estimands.
When we are faced with estimand relevant protocol deviations,
 or inter-current events (e.g.\ rescue
medication) and loss to follow-up {\em etc.}, such assumptions are at
best only partially verifiable from the actual trial data. In such
settings, a primary analysis assumption is made, and then 
the robustness of inferences to a number of secondary sensitivity
assumptions will ideally be explored.  

The assumptions underpinning the primary and
 sensitivity analyses should be as accessible as possible.
This applies not only to assumptions about the typical, or mean,
profile of patients post-deviation, but also to assumptions
about their precision.

In this article, we have introduced the concept of  {\em information-anchoring}---whereby the extent of information loss due to missing
data is held constant across primary and sensitivity analyses. We
believe this facilitates informed inferences and decisions, whatever statistical method is adopted.
 Information-anchoring allows stakeholders to focus on the
assumptions about the mean responses of each patient, or group of patients,
post-deviation, without being concerned as to whether we are injecting information into
or removing information from the analysis (relative to that lost---due to patient deviations---in
the primary analysis).  For example, we believe this provides a good basis for
discussions between regulators and pharmaceutical statisticians: the
former can be reassured the sensitivity analysis is not injecting
information, while the latter can be reassured that the sensitivity
analysis is not discarding information. 

We have differentiated between two different types of sensitivity
analysis: class-1 and class-2.  In class-1 the primary analysis model is
retained in the sensitivity analysis; such sensitivity analyses can be
readily (but need not be) carried out by multiple imputation. 

Controlled MI procedures, which combine a pattern-mixture modelling
approach with MI, naturally fall into this first class. These
include reference-based MI procedures,  which impute missing data
under qualitative 
assumptions for the unobserved data, based on data observed in a
specified reference group. The primary analysis model is retained in
the sensitivity analyses, fitted to each imputed data set and results
combined using Rubin's rules. Consequently  the assumptions of the
primary analysis model are generally inconsistent with the data
generating mechanism postulated by the sensitivity analysis
assumption. Thus the usual justification for Rubin's MI rules does not
hold. Instead, we have identified a new property of
these rules, namely that for a broad class of controlled MI
approaches, including both $\delta$- and reference-based approaches, 
they yield information-anchored inference.
In this regard, a practically
important corollary of our theory is that  the widely used
$\delta$-method (and associated tipping-point analysis) is
information-anchored with fixed $\delta$ adjustment. 

While we believe information-anchored sensitivity analyses provide a
natural starting point, and will often be sufficient, in certain
scenarios it may also be
 desirable to conduct information-negative sensitivity analysis. In
 such analyses a greater loss of information due to post-deviation
 (missing) data is imposed by the analyst in the sensitivity analysis
 relative to the primary analysis. One way to do this is by prior
 elicitation---i.e.\ incorporating a prior distribution on $\delta$---as touched upon in the further comments following Theorem 1 and Section \ref{sec:application}.
The theory in Section \ref{sec:section_4} also shows how a
greater loss of information 
can be imposed in sensitivity analysis via reference-based MI if
required. This is done by  reducing the size of the reference group
used to construct the reference-based imputation models. 

Whatever approach is taken, careful thought needs to be given, and
justification provided, for the additional loss of information being
imposed. As we discussed at the end of Section \ref{sec:application}, the
loss of information with prior elicitation can be substantial.
Often it will be difficult to justify an additional amount of
information loss to impose. 

Conversely, we argue that information-positive sensitivity analysis,
where a lower loss of information due to missing data post-deviation is imposed
in the sensitivity analysis relative to the primary analysis, is
rarely justifiable, if at all. This is because it goes against all our intuition that
missing data means we lose (not gain) information: with information-positive sensitivity analyses, we gain more precise inferences the more data we lose!  

Our approach to determining the appropriate information in sensitivity
analyses (which, as the simple example in the Introduction shows is
under the control of the analyst), contrasts with some recent work.
 \cite{Kaifeng}, \cite{Tang:2017} and \cite{Lu:Pang} each developed alternative implementations
of the reference-based pattern mixture modelling
approach. \cite{Kaifeng} introduced an analytical 
approach for placebo-based (CR) pattern mixture modelling which uses
maximum likelihood and 
the delta method for treatment effect and variance
estimation. \cite{Tang:2017} 
derived different analytical expressions for
reference-based models, also via 
the likelihood-based approach. \cite{Lu:Pang} proposed a Bayesian
analysis for reference-based methods which estimates the treatment
effect and variance from the posterior distribution.   

What these papers have in common is that, in the terminology developed
here, they essentially choose to apply the primary analysis variance estimator 
across the sensitivity analyses.  While this choice  
has a long-run justification, for the reference-based multiple
imputation estimator, as our simulation results in Figure
\ref{fig:simres} show (and we have discussed elsewhere
\cite[]{Carpenter/Roger/Cro/Kenward:2014}), this choice also means
information-positive inferences for reference-based scenarios.
This is a consequence of (i) uncongeniality between the imputation and
analysis model and (ii) the fact that reference-based methods borrow
information from within and across arms. Thus we highlight here that 
if one of these alternative implementations is employed within sensitivity analysis information-positive inference will be obtained. 

What are the implications of this for our approach? Necessarily, the
variance estimate arising from the information-anchored sensitivity
analysis via reference-based multiple imputation  does not have a long-run 
justification for the reference-based multiple imputation point estimate.
However, having determined that the information-anchored variance is
appropriate,  we can readily inflate the long-run variance of
the reference-based multiple imputation estimator
 by adding appropriate random noise. In this way, having chosen
to make our primary and sensitivity analysis information-anchored, we
can derive a corresponding point estimator whose long run variance
is the information-anchored variance. 

If we wish to do this, we can proceed as follows.  
Recall that reference-based methods calculate the
means of the missing values for each patient as linear combinations of 
the estimated treatment means at each time point under randomised arm
MAR. Assume $J$ follow-up visits, and denote these estimated means by the 
$2J \times 1$ column vector $\bmu,$ with estimated covariance matrix $\hat\bV.$
It follows that,  for some $2J\times 1$ column vector $\bL,$ the maximum likelihood
reference-based treatment estimate is given by $\bL^t \bmu,$ 
with associated estimated empirical variance $\hat\sigma^2_{ML}=\bL^t \hat\bV \bL.$
If we denote the information-anchored variance
by $\hat\sigma^2_{IA},$ take a draw from
$N(0,\hat\sigma^2_{IA}-\hat\sigma^2_{ML}),$
add this to the treatment estimate obtained from the reference-based
analysis by MI, this will result in an estimate with the information-anchored variance in a long-run sense. In practice $\hat\sigma^2_{ML}$
could also be estimated using one of the implementations of
\cite{Kaifeng}, \cite{Tang:2017} or \cite{Lu:Pang}.  In applications,
however, we do not think this step is typically worthwhile. 
Note too that with the $\delta$-method $\hat\sigma^2_{IA}$ is well approximated by
$\hat\sigma^2_{ML},$ so it is not necessary.

This article has focused on the analysis of a longitudinal measure
of a continuous outcome. For generalized linear models (GLMs), 
if we perform controlled MI on the linear
predictor scale, then we can apply the theory developed here on the linear predictor
scale. This suggests that for GLMs, controlled MI will be approximately
information-anchored; preliminary simulations support this, and work in this area is
continuing. We note, however, that issues may arise with non-collapsability when
combining the component models in this setting.
For survival data, we need to define the reference-based assumptions. This has been
done in a recent manuscript we have submitted, which also contains simulation results
suggesting promising information-anchoring properties for Rubin's rules in this setting.

When conducting class-1 sensitivity analyses via MI a natural question
might be how many imputations to conduct. As remarked in the proof of
Theorem 1 in Appendix  \ref{app:proof2}, the number of imputations
does not materially affect the information-anchoring performance of
Rubin's variance estimate. Thus we recommend determining the number of
imputations required for primary analysis (under MAR) based on the
required precision; these should estimate the information-anchored
variance with similar precision in sensitivity analysis.  
To establish the number of imputations required to achieve a specific
level of precision under MAR  \cite{Rubin:1987} showed that the relative variance i.e.\ the efficiency of an estimate using only $K$
imputations compared to an infinite number is approximately
$(1+\lambda/K)$, where $\lambda$ is the fraction of missing
information. 
As discussed in \cite{Carpenter/Kenward:2008}, p.\ 86--87,  5--10 imputations is sufficient to get a reasonably accurate answer for most applications.   For more critical inferences, at least 50-100 imputations are recommended (see \citeauthor{Carpenter/Kenward:2013}, (\citeyear{Carpenter/Kenward:2013}), p.\ 54--55).

Of course, to obtain information-anchored analyses Multiple Imputation does not have to be used.  In principle we can perform information-anchored analysis by calculating the variance directly from the information-anchoring formula. 
However, to do this we need to calculate the expected value of the design-variance when
we actually observe data under the sensitivity assumption. When the approach is
used with its full 
flexibility (with different assumptions for different groups of patients)
this is awkward. Multiple imputation provides a much more direct, computationally
general, accessible approach for busy trialists, without the need for
sophisticated one-off programming which is often required to directly fit MNAR pattern-mixture models or other MNAR models. 

In conclusion, we believe that sensitivity analysis via controlled MI
provides an accessible practical approach to exploring the robustness
of inference under the primary assumption to a range of accessible,
contextually plausible
alternative scenarios. It is increasingly being
used in the regulatory world (see, for example, the DIA pages at
\underline{www.missingdata.org.uk}, and the code therein; \cite{jamapsychiatry20152146}, \cite{JCPP:JCPP12443}, \cite{Billingsdc171114}, \cite{jama201720373}, 
 \cite{OKelly/Ratitch:2014} and references therein). Our aim
has been to provide a more formal underpinning. Information-anchoring is a
natural principle for such analysis, and we have shown this is an automatic
consequence of using MI in this setting.

\section{Acknowledgements}
We are grateful to the Associate Editor and two referees whose
comments have lead to a greatly improved manuscript. Suzie Cro was
supported for her PhD by MRC London Hub for Trials Methodology
Research, grant number MC\_EX\_G0800814. 
James Carpenter is supported by the Medical Research Council, grant
numbers MC\_UU\_12023/21 and MC\_UU\_12023/29.

\begin{appendices}
\section{Appendix A: Algorithm for reference-based multiple imputation\label{app:alg}}

For a continuous outcome, the generic algorithm of \cite{Carpenter/Roger/Kenward:2013} can be summarized in full as follows:

\begin{enumerate}
\item Separately for each treatment arm take all the observed data, and assuming MAR, fit a multivariate normal (MVN) distribution with an unstructured mean (i.e. a separate mean for each of the baseline and post-randomisation observation times) and variance covariance matrix using a Bayesian approach with an improper prior for the mean and an uninformative Jeffrey’s prior for the covariance matrix.

\item Draw a mean vector and covariance matrix from the posterior distribution for each treatment arm. Specifically we use the Markov-Chain Monte Carlo (MCMC) method to draw from the appropriate Bayesian posterior, with a sufficient burn-in and update the chain sufficiently in-between to ensure subsequent draws are independent. The sampler is initiated using the Expectation-Maximization (EM) algorithm.

\item Use the draws in step 2 to form the joint distribution of each
  deviating individual's observed and missing outcome data as
  required. This can be done under a range of assumptions, in order to
  explore the robustness of inference about treatment effects. The
  options presented in \cite{Carpenter/Roger/Kenward:2013} that each translate to a relevant assumption are described in Table \ref{tab:options}.

\item Construct the conditional distribution of missing
  (post-deviation) given observed outcome data for
  each individual who deviated, using  their joint distribution formed in
  step 3. Sample their missing post-deviation data from this conditional distributions to create a completed data set.

\item Repeat steps 2--4 $K$ times, resulting in $K$ imputed data sets.
\end{enumerate}

We now describe how step 3 \index{Jump to reference!technical details}
works under `jump to reference'. This leads to a brief presentation of the approach for
the other options. Suppose there
are two arms, active (indexed below by a) and reference (indexed below by r). In step 2, denote the
current draw from the posterior for the
$1+J$ reference arm means and variance-covariance matrix
by $\mu_{r,0},\dots\mu_{r,J},$ and $\bSigma_r.$
Use the subscript $a$ for the
corresponding draws from the other arm in question (which will
depend on the arm chosen as reference for the analysis at hand).

Under `jump to reference', suppose patient $i$ is not randomised to the reference
arm and their last observation, prior to deviating, is at
 time $d_i,$ $d_i\in (1,\dots,J-1).$
The joint distribution of their observed and post-withdrawal
outcomes is multivariate normal with mean
\[
\tilde\bmu_i=(\mu_{a,0},\dots,\mu_{a,d_i},\mu_{r,d_i+1},\dots,\mu_{r,J})^T;
\]
that is post-deviation they `jump to reference'.

We construct the new covariance matrix for these observations as
follows. 
Denote the covariance matrices from the reference arm (without deviation)
and
the other arm in question (without deviation), partitioned at time
$d_i$ according to the pre- and
post-deviation measurements, by:
\[
\mbox{Reference}\;\;
\bSigma_r = \mat{cc}{
\bR_{11} & \bR_{12} \\
\bR_{21} & \bR_{22}
}\;\;
\mbox{and other arm:}\;\;
\bSigma_a = \mat{cc}{
\bA_{11} & \bA_{12} \\
\bA_{21} & \bA_{22}
}.
\]

We want the new covariance matrix, $\bSigma$ 
 say, to match that from the active arm for the pre-deviation measurements,
and the reference arm for the \emph{conditional} components for the post-deviation
given the pre-deviation measurements. This also guarantees positive definiteness of the new matrix,
since $\bSigma_r$ and $\bSigma_a$  are positive definite.
That is, we want
\[
 \bSigma = \mat{cc}{
 \bSigma_{11} &  \bSigma_{12} \\
\bSigma_{21} &  \bSigma_{22}
},
\]
subject to the constraints
\begin{align*}
\bSigma_{11} &= \bA_{11},\\
 \bSigma_{21}\bSigma_{11}^{-1} &= \bR_{21}\bR_{11}^{-1},\\
 \bSigma_{22} - \bSigma_{21} \bSigma_{11}^{-1}\bSigma_{12} &= \bR_{22} - \bR_{21}\bR_{11}^{-1}\bR_{12}.
\end{align*}

The solution is:
\begin{align*}
&& \bSigma_{11} &= \bA_{11},\\
&& \bSigma_{21} &= \bR_{21}\bR_{11}^{-1}\bA_{11},\\
&& \bSigma_{22} &= \bR_{22}- \bR_{21}\bR_{11}^{-1}(\bR_{11}-\bA_{11})\bR_{11}^{-1}\bR_{12}.
\end{align*}

Under `jump to reference' we have now specified the joint distribution for a patient's
pre- and post-deviation outcomes, when deviation is at time $d_i.$ 
This is what we require for step 4. For `copy increments in reference' we use the same $ \bSigma$ as for `jump to reference'
but now
\begin{eqnarray}
\bmu_i&=\left\{\mu_{a,0},\dots,\mu_{a,d_i-1},\mu_{a,d_i},\mu_{a,d_i}+(\mu_{r,d_{i}+1}-\mu_{r,d_{i}}),\right. \nonumber \\
& \;\;\left.\mu_{a,d_i}+(\mu_{r,d_{i}+2}
-\mu_{r,d_i}), \ldots\right\}^T. \nonumber
\end{eqnarray}
For `last mean carried forward', $ \bSigma$ equals the covariance
matrix from the randomisation arm. The important change
is the way we put together
$\bmu$. Thus, for patient $i$ in arm $a$ under `last mean carried
forward',
\[
\bmu_i=(\mu_{a,0},\dots,\mu_{a,d_i-1},\mu_{a,d_i},\mu_{a,d_i},\dots...)^T;
\quad \bSigma=\bSigma_a.
\]
Finally for `copy reference' the mean and covariance both come from
the reference (typically, but not necessarily, control) arm,
irrespective of deviation time. A SAS macro implementing this approach can be downloaded from, \\ \underline{www.missingdata.org.uk} (\citeauthor{macros:2012}, \citeyear{macros:2012})
 and Stata software from \\ \underline{https://ideas.repec.org/c/boc/bocode/s457983.html} (\citeauthor{Cro15}, \citeyear{Cro15}; \citeauthor{Cro/Morris/Kenward/Carpenter:2016}, \citeyear{Cro/Morris/Kenward/Carpenter:2016}).

\section{Appendix B: Proofs}

\subsection{Proof of Proposition 1 \label{app:proof1}}

Here we outline the argument for Proposition 1. Consider the baseline (time 1) and $J-1$ follow-up setting where $Y_{z,i,j}$  denotes the continuous outcome measure for patient $i$ in arm $z$ ($z=a$ indicates active arm allocation and $z=r$ reference arm allocation) at time $j$ for $i=1,...,n$ and $j=1,...,J$. $n_{d,j}$
patients deviate at time $j$ in a monotone fashion, for $j>1$ such that
$n_d=\sum_{j=2}^{J}n_{d,j}$. Interest lies in the unadjusted mean treatment group 
difference at time $J$. Conditioning on $n_{d,j}$ for $j>1$, the expected value of the treatment estimate at time $J$ when the post-deviation data can be observed is,
\[\left(\frac{n_o}{n}\mu_{a,J} + \sum_{j=2}^{J}\frac{n_{d,j}}{n}\mu_{d,j,J}\right) - \mu_{r,1}.
\]
The variance of this estimate is calculated using the usual sample variance formula as,
\[
\frac{\frac{1}{n-1}\sum_{i=1}^{n}\left(Y_{r,i,J}-\bar{Y}_{r,J}\right)^2}{n} + \frac{\frac{1}{n-1}\sum_{i=1}^{n}\left(Y_{a,i,J}-\frac{n_o}{n}\bar{Y}_{a,J,o}-\sum_{j=2}^{J}\frac{n_{d,j}}{n}\bar{Y}_{a,J,d,j}\right)^2}{n}
\]
where $\bar{Y}_{r,J}=\frac{1}{n}\sum_{i=1}^{n}Y_{r,i,J}$, $\bar{Y}_{a,J,o}=\frac{1}{n_o}\sum_{i\in o}Y_{a,i,J}$ and $\bar{Y}_{a,J,d,j}=\frac{1}{n_{d,j}}\sum_{i\in d,j}Y_{a,i,J}$ for $j=2,...,J$. When expanding this expression and letting $(n-1)\rightarrow n$ this has expected value,
\[
E\left[\vsfulhat\right]=\frac{\sigma_{J,J}^2}{n}+\frac{\sigma_{J,J}^2}{n}+\sum_{j=2}^{J}\frac{n_on_{d,j}\Delta_{d,j}^2}{n^3} + \sum_{p=2}^{J}\sum_{q=2}^{\overset{J}{q\neq p}}\frac{n_{d,p}n_{d,q}\Delta_{d,p,q}^2}{n^3}
\]
where $\Delta_{d,j}=\mu_{a,J}-\mu_{d,j,J},\Delta_{d,p,q} = \mu_{d,p,J}-\mu_{d,q,J}$, $\mu_{d,j,J}$ is the mean proposed under the controlled scenario at time $J$, for patients who deviate at time $j$ and $\mu_{d,p,J}$ and $\mu_{d,q,J}$ are the means proposed under the controlled scenario at time $J$, for patients who deviate at times $p$ and $q$ for $p=2,...,J$ and $q=2,...,J$. For the $\delta$-method of MI where imputed values at final time $J$ are edited by $(J+1-j)\delta$, for patients who deviate at time $j$, we replace $\Delta_{d,j}=\mu_{a,J}-\mu_{d,j,J}=(J+1-j)\delta$ and
$\Delta_{d,p,q}=\mu_{d,p,J}-\mu_{d,q,J}=(J+1-p)\delta-(J+1-q)\delta$.

\subsection{Proof of Theorem 1 \label{app:proof2}}

Let $\mathcal{D}$ and $\mathcal{O}$ define the sets of indices for the patients who
do and do not deviate in the active arm respectively. Further let $\mathcal{DJ}$ denote
the set of indices for deviating patients who deviate at time $j$, so that the total number of deviating patients in the active arm $n_d=\sum_{j=2}^{J}n_{d,j}$. The follow-up outcome data at the final time point for the reference
patients are contained in the vector $\mathbf{Y}_{r,J}=\left(Y_{r,1,J},...Y_{r,n,J}\right)^T$. 
The final visit outcome 
data for the observed non-deviating active patients are contained
 in the vector  $\mathbf{Y}_{a,J,o}=\left\{Y_{a,i,J};i\in \mathcal{O}\right\}^T$.

We suppose that each deviating patient has two potential  outcomes at time $J$:
 the one that would occur if they remain on active treatment post-deviation (primary on-treatment data model, indexed below with the subscript $_P$)
 and the other that would occur under the controlled sensitivity  scenario data model (indexed below with the subscript $_S$). The potentially observable primary on-treatment data for the $n_d$ deviating
 patients at time $J$ are contained in the vector $\mathbf{Y}_{a,J,P,d}$
 and the alternative  outcome data under the sensitivity scenario in the vector
 $\mathbf{Y}_{a,J,S,d}$. Define \\  $\mathbf{Y}=\left(\mathbf{Y}_{r,J}, \mathbf{Y}_{a,J,o},\bY_{a,J,P,d},\bY_{a,J,S,d} \right)^T$ as the collection of observed
and potentially observable outcome data at time $J$,
which has dimensions $\left[(n+n_{o}+2n_d)\times1\right]$. 

For each deviating patient we can only observe one of the potential outcomes, 
either primary on-treatment or under the sensitivity scenario. Consider two \\ $\left[\left(n+n_o+2n_d\right)\times\left(n+n_o+2n_d\right)\right]$ matrices,
 $\mathbf{D_{P}}$ and $\mathbf{D_{S}}$ of 0's and 1's such that  
$\mathbf{D_{P}Y}$ gives the $\left[\left(n+n_o+2n_d\right)\times 1 \right]$ on-treatment (primary) data and $\mathbf{D_{S}Y}$ \\ gives the $\left[\left(n+n_o+2n_d\right)\times 1 \right]$ sensitivity scenario data
at time $J$.

Let $\mathbf{a}$ be a $\left[(n+n_{o}+2n_d)\times1\right]$
vector such that $\mathbf{a^TD_{P}Y}$ returns the primary on-treatment treatment 
estimate and $\mathbf{a^TD_{S}Y}$ returns the sensitivity scenario treatment estimate. 
When the deviating patients experience primary on-treatment behaviour post-deviation and are fully observed the expectation of the variance
 of the primary on-treatment estimand can be expressed as,
\begin{equation}
\text{E}\left[\vpfulhat\right]=\text{E}\left[\right.V\left.\left(\mathbf{a^TD_{P}Y}\right)\right]=\text{E}\left[\mathbf{a^TD_{P}}\right.V\left.\mathbf{(Y)D_{P}^Ta}\right]=\mathbf{a^TD_{P}\Sigma D_{P}^Ta}. \label{eq:varone}
\end{equation}
Under the  conditions of Proposition 1 and using Corollary 1 and 2, the variance estimator for the sensitivity estimand where post-deviation data are fully observed can be expressed as,
\begin{equation}
E\left[\vsfulhat\right]=\mathbf{a^TD_{P}\Sigma D_{P}^Ta}+O(n^{-2}).\label{eq:vartwo}
\end{equation}
We now suppose that post-deviation data are unobserved, i.e.\ the potentially
observable primary on-treatment and sensitivity scenario entries in $\mathbf{Y}$ are missing 
for the $n_d$ active patients. We alternatively multiply impute these outcomes, 
using primary on-treatment (MAR) imputation and imputation under the sensitivity scenario. This gives $K$ `complete' 
data samples $\mathbf{Y}_k$, of size $\left[(n+n_{o}+2n_d)\times1\right]$.
For this we need appropriate imputation distributions for each missing data pattern
under each scenario, with suitable posteriors for the included parameters.

Under our primary on-treatment assumption (MAR), the imputation model for patients deviating at time $j$, for each $j>1$ is formed from the regression of $\mathbf{Y}_{a,J,o}$ on $\mathbf{P}_{a,o,j}$ where $\mathbf{P}_{a,o,j}$ is the design matrix for the imputation model, 
which contains the values of the $1,...,j-1$ outcomes and covariates included 
in the imputation model (excluding treatment) for the $n_{o}$ observed active patients, along with a vector of 1's to include an intercept in the model. This is appropriate since we are not imputing any interim missing outcomes here. We only consider monotone missing data patterns. We are interested in the treatment effect at time $J$. As described by Carpenter and Kenward (2013, p.\ 77--78), under MAR, each of the regressions will be validly estimated from those observed in the data set. The parameter estimates for the primary on-treatment (MAR) imputation model for the $n_{d,j}$ 
patients missing outcomes $j$ to $J$ for each $j>1$ are found as 
$\hat{\bbe}_{primary,j}= (\bP_{a,o,j}^T\bP_{a,o,j})^{-1}\bP_{a,o,j}^T\bY_{a,J,o}$ 
with assumed known covariance matrix $\mathbf{V}_{primary,j}=(\bP_{a,o,j}^T\bP_{a,o,j})^{-1}\sigma_j^2$.

We assume the large sample posterior for the parameter estimates for the primary on-treatment imputation model, denoted $\bbeh_{primary,j}$, is 
normal and centered on the ML estimator $\bbeh_{primary,j}$ 
with covariance matrix $\mathbf{V}_{primary,j}$. That is,
\[
\bbeh_{primary,j}|Y_{a,J,o}\sim N(\bbeh_{primary,j};\bV_{primary,j}).
\]
The primary on-treatment imputation model for active
patient $i$ deviating at time $j$, for each $j>1$ and imputation $k$ can therefore be expressed as,
\[
\tilde{Y}_{a,i,J,k}|\mathbf{Y}_{a,J,o} = \mathbf{P}_{a,d,j,i}\left[\bbeh_{primary,j}+\bb_{primary,j,k}\right]+e_{i,j,k} \mbox{ for } i\in{\left\{\mathcal{DJ}\right\}},
\]
where, $\mathbf{b}_{primary,j,k}\sim N(0,\mathbf{V}_{a,o,j}),e_{i,j,k}\sim N(0,\sigma_j^2)$ and $\mathbf{P}_{a,d,j,i}$ contains the values of the $1,...,j-1$ outcomes and covariates included 
in the imputation model (excluding treatment, plus a 1 for the intercept) for each deviating active 
patient $i$, who deviates at time $j$.

For sensitivity analysis we conduct imputation under the proposed sensitivity scenario and assume the large sample posterior for the 
imputation parameters for the $n_{d,j}$ patients missing outcomes $j$ to $J$ for each $j>1$, 
$\hat{\bbe}_{sensitivity,j}$ is normal and centered on the ML 
estimator $\bbeh_{sensitivity,j}$ with known covariance matrix 
$\mathbf{V}_{sensitivity,j}$, that is for each $j>1$,
\[
\bbeh_{sensitivity,j}|\bY_{sensitivity,J}\sim N(\bbeh_{sensitivity,j};\bV_{sensitivity,j}),
\]
where $\mathbf{Y}_{sensitivity,J}$ consists of the relevant observed outcome data under 
the particular sensitivity scenario setting of interest. The imputation 
model used in the sensitivity analysis for active patient $i$ deviating following time $j$, for each $j>1$ and imputation $k$ can therefore be expressed as,
\[
\breve{Y}_{a,i,J,k}|\mathbf{Y}_{sensitivity,J} = \mathbf{P}_{a,d,j,i}\left[\bbeh_{sensitivity,j}+\bb_{sensitivity,j,k}\right]+e_{i,j,k} \mbox{ for } i\in{\left\{\mathcal{DJ}\right\}},
\]
where, $\mathbf{b}_{sensitivity,j,k}\sim N(0,\mathbf{V}_{sensitivity,j}) \mbox{ and } e_{i,j,k}\sim N(0,\sigma_j^2)$. Under the assumption of equal variance-covariance matrix of baseline and follow-up by treatment arm we consequently assume the same variance for the residuals in the primary and sensitivity imputation models for patients deviating at the same time $j$, for each $j>1$. 

We are interested in imputation inference for,  
$\frac{1}{K}\sum_{k=1}^K\mathbf{a^TD_{P}Y}_k$ or  $\frac{1}{K}\sum_{k=1}^K\mathbf{a^TD_{S}Y}_k$. Letting the number of imputations, $K\rightarrow\infty$, 
the variance of our MI treatment estimate as estimated by Rubin's rules is,
$\vpmihat=\wprimary+\bprimary$ or $\vsmihat=\wsen+\bsen$ where under the conditions required  in the proposition,\\ $E\left[\wprimary\right]=E\left[\frac{1}{K}\sum_{k=1}^K\mathbf{a^TD_{P}\hat{\Sigma}_kD_{P}^Ta}\right]\rightarrow\mathbf{a^TD_{P}\Sigma D_{P}^Ta}$ and, \\ $E\left[\wsen\right]=E\left[\frac{1}{K}\sum_{k=1}^K\mathbf{a^TD_{S}\hat{\Sigma}_kD_{S}^Ta}\right]\rightarrow\mathbf{a^TD_{P}\Sigma D_{P}^Ta}+O(n^{-2})$.

Under primary (on-treatment) imputation,
\[
\bprimary=\frac{1}{K-1}\sum_{k=1}^K\left[\sum_{j=2}^{J}\pi_{d,j}\left(\bar{e}_{j,k}-\bar{e}_j\right)+\pi_{d,j}\left(\mathbf{\bar{P}}_{a,d,j}\mathbf{{b}}_{primary,j,k}-\mathbf{\bar{P}}_{a,d,j}\mathbf{\bar{b}}_{primary,j}\right)\right]^2
\]

where $\bar{e}_{j,k}=\frac{1}{n_{d,j}}\sum_{i\in \mathcal{DJ}} e_{i,j,k}$, $\bar{e}_j=\frac{1}{K}\sum_{k=1}^K \bar{e}_{j,k}, \mathbf{\bar{P}}_{a,d,j}= \frac{1}{n_{d,j}}\sum_{i\in \mathcal{DJ}} \mathbf{P}_{a,d,j,i}$ and \\ $\mathbf{\bar{b}}_{primary,j}=\frac{1}{K}\sum_{k=1}^K \mathbf{{b}}_{primary,j,k}$.  Which has expectation,
\begin{eqnarray}
E\left[\bprimary\right] & =\sum_{j=2}^{J} \pi_{d,j}^2\left[\frac{\sigma_j^2+n_{d,j}\mathbf{\bar{P}}_{a,d,j}\mathbf{V}_{primary,j}\mathbf{\bar{P}}_{a,d,j}^T}{n_{d,j}}\right]. \nonumber
\end{eqnarray}
When imputation is conducted under the sensitivity scenario,
\[
\bsen=\frac{1}{K-1}\sum_{k=1}^K\left[\sum_{j=2}^{J}\pi_{d,j}\left(\bar{e}_{j,k}-\bar{e}_j\right)+\pi_{d,j}\left(\mathbf{\bar{P}}_{a,d,j}\mathbf{{b}}_{sensitivity,j,k}-\mathbf{\bar{P}}_{a,d,j}\mathbf{\bar{b}}_{sensitivity,j}\right)\right]^2,
\]
where $\mathbf{\bar{b}}_{sensitivity,j}=\frac{1}{K}\sum_{k=1}^K \mathbf{{b}}_{sensitivity,j,k}$. Which has expectation,
\begin{eqnarray}
E\left[\bsen\right] & =\sum_{j=2}^{J} \pi_{d,j}^2\left[\frac{\sigma_j^2+n_{d,j}\mathbf{\bar{P}}_{a,d,j}\mathbf{V}_{sensitivity,j}\mathbf{\bar{P}}_{a,d,j}^T}{n_{d,j}}\right]. \nonumber
\end{eqnarray}
The information-anchored variance can be expressed as,
\[
E\left[\vanchat\right] = \frac{E\left[\vsfulhat\right]\left(E\left[\wprimary\right]+E\left[\bprimary\right]\right)}{E\left[\vpfulhat\right]}=E\left[\vsfulhat\right]\left[1+\frac{E\left[\bprimary\right]}{E\left[\wprimary\right]}\right].
\]
Since $E\left[\wprimary\right]=E\left[\vpfulhat\right]$ and using \eqref{eq:varone} and \eqref{eq:vartwo} that is,
\begin{eqnarray}
E\left[\vanchat\right]  &= \mathbf{a^TD_{P}\Sigma D_{P}^Ta}+O(n^{-2}) +\frac{E\left[\bprimary\right]}{E\left[\wprimary\right]}\left[\mathbf{a^TD_{P}\Sigma D_{P}^Ta}+O(n^{-2})\right] \nonumber  \\
 & = \mathbf{a^TD_{P}\Sigma D_{P}^Ta}+O(n^{-2}) + E\left[\bprimary\right]+\frac{E\left[\bprimary\right]}{E\left[\wprimary\right]}O(n^{-2}). \nonumber
\end{eqnarray}
If Rubin's rules are information-anchoring and preserve the information loss in the primary analysis under MAR then the following holds,
\[
E\left[\wsen\right]+E\left[\bsen\right]\approx \mathbf{a^TD_{P}\Sigma D_{P}^Ta}+O(n^{-2}) + E\left[\bprimary\right]+\frac{E\left[\bprimary\right]}{E\left[\wprimary\right]}O(n^{-2}).
\]
That is,
\begin{eqnarray}
\mathbf{a^TD_{P}\Sigma D_{P}^Ta}+O(n^{-2})+E\left[\bsen\right]\approx & \mathbf{a^TD_{P}\Sigma D_{P}^Ta}+O(n^{-2}) \nonumber \\
& + E\left[\bprimary\right]+\frac{E\left[\bprimary\right]}{E\left[\wprimary\right]}O(n^{-2}). \nonumber
\end{eqnarray}

After simplification and rearrangement this becomes,
\begin{eqnarray}
0 \approx E\left[\bprimary\right] - E\left[\bsen\right] + \frac{E\left[\bprimary\right]}{E\left[\wprimary\right]}\left[O(n^{-2})\right].  \nonumber
\end{eqnarray}
Which is,
\begin{eqnarray}
0 \approx \sum_{j=2}^{J} \left[\pi_{d,j}^2\mathbf{\bar{P}}_{a,d,j}\left(\mathbf{V}_{primary,j}-\mathbf{V}_{sensitivity,j}\right)\mathbf{\bar{P}}_{a,d,j}^T\right] + \frac{E\left[\bprimary\right]}{E\left[\wprimary\right]}\left[O(n^{-2})\right] \nonumber
\end{eqnarray}

This gives the required result in the longitudinal trial setting with monotone missingness in the active treatment arm with $K=\infty$. In practice $K\neq\infty$, however the information-anchoring approximation results will still hold for finite $K$. For finite $K$ the variance of our MI treatment estimate as estimated by Rubin's rules is,
$\vpmihat=\wprimary+\left(1+\frac{1}{K}\right)\bprimary$ or $\vsmihat=\wsen+\left(1+\frac{1}{K}\right)\bsen$. We will therefore have additional terms in the difference between Rubin's variance estimator and the ideal information-anchored variance, but these will also be very small. They will be the same order of the terms already presented multiplied by $K^{-1}$, hence indeed smaller. Thus following the reasons discussed in the main text the approximation remains with finite $K$.

We note that when we relax the equal variance by trial arm assumption, we can no longer assume the variance of the residuals
in the primary de jure imputation model for patients with missingness pattern $j$ matches the variance of
the residuals in the sensitivity de facto imputation model for patients with missingness pattern $j$, for each
missing data pattern $j$.

In this setting we denote the variance of the residuals in the primary on-treatment imputation model for patients missing outcomes $j,...,J$  as $\sigma_{P,j}^2$ and in the sensitivity imputation model as $\sigma_{S,j}^2$ for $j>1$. Then the information-anchoring performance of Rubin's MI variance estimator is driven by,
\begin{eqnarray}
0 \approx \sum_{j=2}^{J} \pi_{d,j}^2\left[\frac{\sigma^2_{P,j}-\sigma^2_{S,j}}{n_{d,j}}+\mathbf{\bar{P}}_{a,d,j}\left(\mathbf{V}_{primary,j}-\mathbf{V}_{sensitivity,j}\right)\mathbf{\bar{P}}_{a,d,j}^T\right] + \frac{E\left[\bprimary\right]}{E\left[\wprimary\right]}\left[O(n^{-2})\right].  \nonumber
\end{eqnarray}

The additional components in the difference between Rubin's variance and the ideal information-anchored variance are driven by the degree of difference in the variance structure of the data by trial arm for each missingness pattern. Since the variance structure is not likely to differ too markedly by trial arm for each missingness pattern, and these extra components are each multiplied by $\pi_{d,j}^2/n_{d,j}$, the overall impact will in practice be relatively small.

\newpage

\section{Appendix C: Further simulation results\label{app:sims}}

\begin{figure}[h]
\centering
\makebox{\vspace*{-0cm}\hspace*{-.1cm}\includegraphics[width=11.7cm]{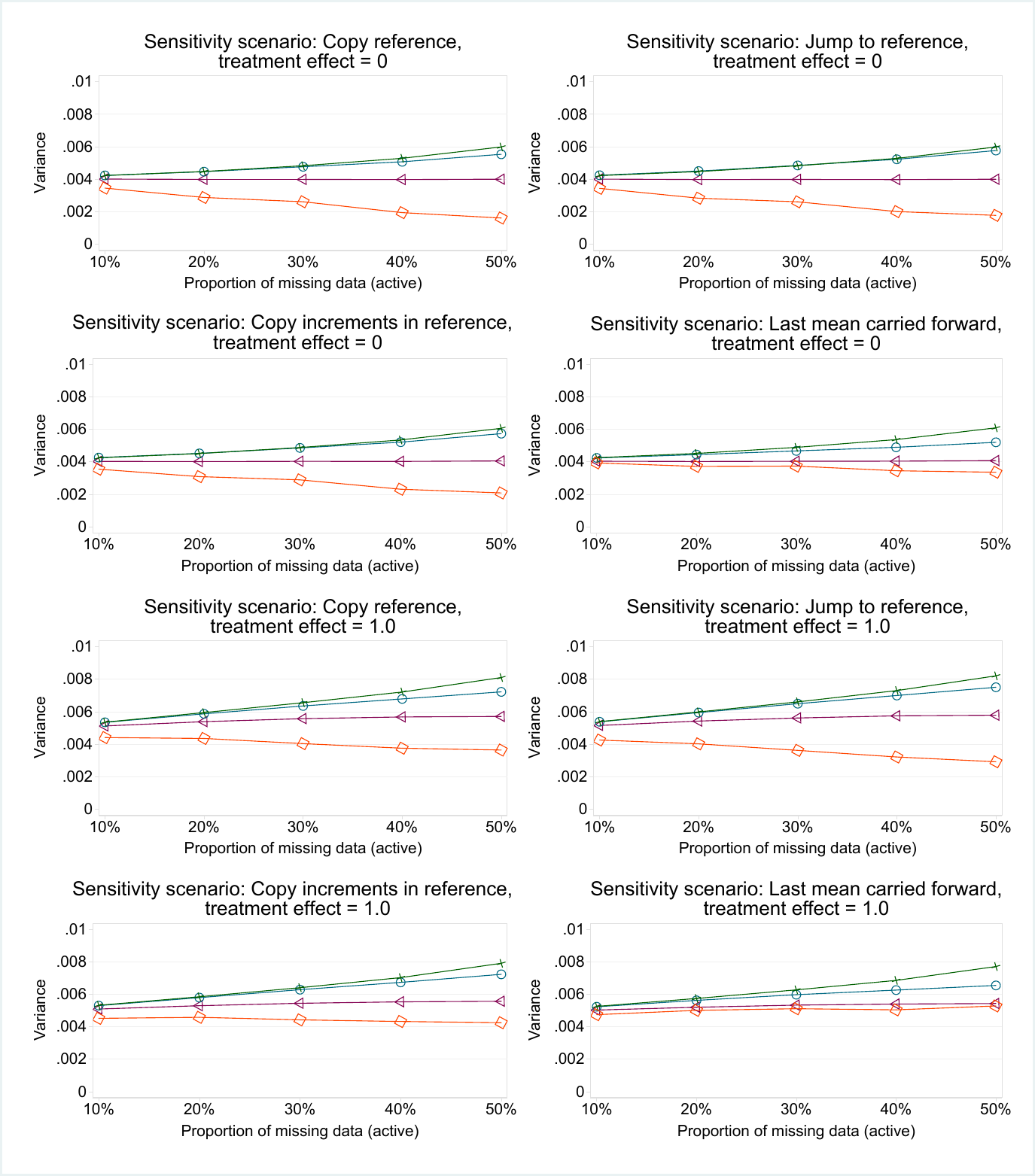}} 
\vspace*{-.1cm}
\caption{\label{fig:simres2}Simulation results: for each sensitivity
  scenario, as the proportion of
active arm deviations increases, each panel shows the evolution of the
mean estimate of the visit 3 treatment effect variance (over 1000 simulations) calculated in
four ways:  (i) $-\!\!\circ\!\!-$ Rubin's MI variance, from
reference-/$\delta$-based MI;  (ii)  $-\!\!\times\!\!-$
Information-anchored variance ($\hat\E[\hat{V}_{\textnormal{\tiny anchored}}]$);  (iii) $-\!\!\diamond\!\!-$ applying primary analysis
variance estimator in sensitivity scenarios; (iv) $-\!\!\triangleleft\!\!-$ Variance when post-deviation data is alternatively fully observed under the given scenario ($\hat\E[\vsfulhat]$).}
\end{figure}

\end{appendices}
\clearpage

\bibliographystyle{JRSS}

\bibliography{Information_anchoring}

\end{document}